\documentclass[%
 reprint,
 amsmath,amssymb,
 aps,
]{revtex4-1}
\usepackage[utf8]{inputenc}
\usepackage{graphicx}
\usepackage{slashbox}
\usepackage{sidecap}
\usepackage{amsthm,amsmath}
\usepackage{framed}
\usepackage[font={footnotesize,it}]{caption}
\usepackage{color}
\usepackage{marvosym}
\usepackage{amssymb}
\theoremstyle{definition} 

\newcommand{\vast}{\bBigg@{4}}
\newcommand{\Vast}{\bBigg@{5}}

\newcommand{\dcell}[2][c]{%
  \begin{tabular}[#1]{@{}c@{}}#2\end{tabular}}

\newcommand{\eq}[1]{\,\begin{equation}
                   #1 
                   \end{equation}
}

\newcommand{\morabba}[1]{\,\begin{flushright}
 \Rectsteel \\
\end{flushright}}

\newcommand{\eqq}[2]{\,\begin{equation} \label{#2}
                   #1 
                   \end{equation}
}

\usepackage{setspace}

\newcommand{\al}[1]{\,\begin{align}
                   #1 
                   \end{align}
}

\usepackage[top=24mm, bottom=25mm, left=22mm, right=20mm]{geometry}

\begin{document}
\preprint{APS/123-QED}

\title{Network Growth with Arbitrary Initial Conditions: Degree Dynamics for Uniform and Preferential Attachment}

\author{Babak Fotouhi and Michael G. Rabbat \\
Department of Electrical and Computer Engineering\\
McGill University, Montr\'eal, Qu\'ebec, Canada\\
Email:\texttt{ babak.fotouhi@mail.mcgill.ca, michael.rabbat@mcgill.ca}\\}


\begin{abstract}
This paper provides time-dependent  expressions for the expected degree distribution of a given network that is subject to growth, as a function of time. We consider both uniform attachment, where incoming nodes form links to existing nodes selected uniformly at random, and preferential attachment, when    probabilities are assigned proportional to the degrees of the existing nodes. We   consider the cases of single and multiple links being formed by each newly-introduced node.  The initial conditions are arbitrary, that is, the solution depends on the degree distribution of the initial graph which is the substrate of the growth. Previous work in the literature focuses on the asymptotic state, that is, when the number of nodes added to the initial graph tends to infinity, rendering the effect of the initial graph negligible. Our contribution provides a solution for the expected degree distribution as a function of time, for arbitrary initial condition. Previous results match our results in the asymptotic limit. The results are discrete in the degree domain, and continuous in the time domain, where the addition of new nodes to the graph are approximated by a continuous arrival rate.

\end{abstract}
\maketitle

\section{Introduction} \label{sec:intro}
The complex network literature spans various strands of research such as sociology~\cite{soci1, soci2,soci3},  economics~\cite{econ1, econ2}, computer science~\cite{cs1, cs2, cs3}, marketing~\cite{mrk1, mrk2, mrk3, mrk4}, epidemiology~\cite{epid1, epid2}, genetics~\cite{gen1}, and bibliometrics~\cite{bibl1}. These domains aim to extract macro-scale behavior from given micro-scale interactions.. 

The structure of the underlying graph, which connects the agents and consequently regulates their interactions, is necessary for studying the dynamism of various phenomena, such as flow of information (news, rumors, trends, etc.) in the society~\cite{diffus1, diffus2},  resilience against node or link failures (for the internet, it means survival of the system if certain nodes are shut down)~\cite{resil1, resil2}, pace of diffusion of a contagious  disease throughout a population and also optimal immunization strategies~\cite{immun1, immun2, immun3}, the effect of the network structure among actors on their chance of winning awards~\cite{oscar1}, to name a few. Models have been proposed to emulate different  structural 
properties observed in real life graphs~\cite{price1, price2, ER1,ER2, bollobas1,Bollobas2,watts1,watts2, barabSF1,barabSF2}. 

In many applications, such as the world wide web~\cite{WWWSF1, WWWSF2} and scientific collaborations~\cite{collabSF1}, networks are dynamic, that is, subject to growth. This provides motivation to view the  problem of network formation dynamically. In this formulation, nodes are introduced successively, and they select from existing nodes whom to attach to. It mimics, for example, the mechanism by which new papers cite existing ones. \cite{barabSF2} takes this approach and introduces the \emph{preferential attachment} mechanism, which is explained below. Also, in~\cite{rednerM1, rednerM2, rednerM3, rednerM4} the problem is tackled by the conventional techniques  of polymer physics. Both of these approaches employ approximations to solve the problem. In what follows, we go over these approximations and the corresponding results.

\subsection{Previous Work: Network Growth}
In the linear preferential attachment scheme  introduced in~\cite{barabSF2}, the growth mechanism is as follows. The growth process starts 
with $N(0)$ nodes. Then, nodes are introduced one per unit time. Each node picks $\beta$ existing nodes to link to, with probabilities assigned to them proportional to their degrees. This means that an existing node with a higher degree will be more likely to attract the newly-introduced node.  Denote the degree distribution of the graph when the total number of nodes is $t$ by $p_k(t)$. 
Their result can be expresses as follows: 
\eqq{
\lim_{t\rightarrow \infty} p_k(t) \sim k^{-3}.
}{barab_res_1}

The analysis is done within the mean-field simplification and the solution is valid in the asymptotic case of $t\rightarrow \infty$.
In~\cite{Bollobas2}, this result is ameliorated  by reformulating the problem more rigorously. Denote  by $\beta$ the number of links that each newly-born node emanates to the existing nodes. The network growth process starts from a $4-$cycle. Let $p_k(N)$ be defined as above.  Also, define: 
\eqq{
\xi_{\beta,k} \stackrel{\text{def}}{=} \frac{2(\beta)(\beta+1)}{k(k+1)(k+2)}.
}{bol_res_1}
In Theorem 1 in~\cite{Bollobas2}, it is shown that  in the limit $t\rightarrow \infty$, for any positive $\epsilon$ and for $0\leq k \leq N^{1/5}$, the following holds: 
\eq{
(1-\epsilon) \xi_{\beta,k} \leq p_k(N) \leq (1+\epsilon) \xi_{\beta,k}.
}
Note that the expression in~\eqref{bol_res_1} agrees with~\eqref{barab_res_1} for large values of $k$. 

The problem is also closely related to the so-called Polya's urn problem~\cite{polya1} in combinatorics. Given a finite number of bins, additional balls arrive one at a time. With a given probability, a new bin is created for the new ball. The ball otherwise joins an existing bin. It picks the destination bin with probabilities dependent on the existing number of balls within the bins. In~\cite{polya1}, the case where probabilities are proportional to $m^{\gamma}$ is solved. The case of  $m=\gamma$ is akin to the linear preferential attachment scheme mentioned above. 

A novel way to tackle the problem was presented in~\cite{rednerM1, rednerM2, rednerM3, rednerM4, rednerM5} by employing the master equation approach which authors borrow from polymer physics. The result of~\eqref{bol_res_1} for the case of $\beta=1$ has been obtained using this approach (equation (2) in\cite{rednerM1}, (5) in~\cite{rednerM5}, (2) in~\cite{rednerM2} and (2) in~\cite{rednerM4}). For a treatment of finite size effects (when $N$ is not infinitely large) with primary focus on nodes with degree ${k_{\textnormal{max}}\approx \sqrt{N}}$, see~\cite{rednerM4}. In~\cite{drog_net},   the generating function approach has been used to solve the master equation and  the asymptotic distribution~\eqref{bol_res_1} for~$t \rightarrow \infty$ has been recovered, and    the asymptotic degree distribution up to the leading order of~$1/t$  has been obtained (for the initial condition of a single node with a specified number of incoming links from outside the network, since in~\cite{drog_net}, directed links can originate from unspecified sources, even from outside the network), in the form of~$p(k,s,t)$, as a function of time~$t$ and time of birth~$s$, which is the time at which each node is introduced to the network.  In the present paper, we seek ~$p_k(t)$  for arbitrary times, and links are necessarily emanated from the newly introduced nodes at each timestep, and also the links are undirected.

In~\cite{rednerM1}, the uniform attachment scheme is also examined. This means that, new nodes attach to existing nodes with equal probabilities, regardless of their degrees. If we start from a single node at the outset, the resulting graph is called a \emph{Random Recursive Tree (RRT)}. The result presented in~\cite{rednerM1} for the asymptotic degree distribution of RRTs is as follows: 
\eqq{
\lim_{t \rightarrow \infty} p_k(t) = \frac{1}{2^k}.
}{red_res_2}
The same result is also presented in Theorem 1 in\cite{RRTmath1} and equation (49) in~\cite{RRTmath2} following a combinatorial approach.

\subsection{Time-dependent Solution, Motivation}
Previous work has been primarily revolved around the asymptotic degree distribution, that is, when the number of nodes tends to infinity. Also, in some case, further simplification is acquired by limiting the range of degrees. For long times, the effect of the initial graph is neglected. In this contribution, we start from an initial graph with known degree distribution $n_k$. We solve for the expected degree distribution at time $t$. We consider both uniform and linear preferential attachment (eventuating in a scale-free graph in the long run). 
The time-dependent solution, first develops intuition about the growth process, and the path that the system undergoes until it reaches the steady state. More importantly, the effect of the initial conditions is taken into account. Different substrates reach the equilibrium approximation of the degree distribution which is at hand, with different paces. The time-dependent solution illuminates the effect of the initial condition on the  accuracy of the above-mentioned approximations.

Equipped with the time-dependent solution, one can also examine the short-time behavior, in marked contrast with the convention, which limits the solution to the long-time behavior. As an example of how  the need for extracting the short-time growth of an existing graph is elicited in realistic applications, consider the network of supporters in a political campaign. Nodes are fanatics who absorb new people into the campaign, causing the network to expand throughout the potential electorate. The change in the network of followers in one day is not substantial compared to the existing size of the network. As another example, consider the social network within a country, with a small number of immigrants joining and enlarging the network. The number of immigrants typically constitutes a small fraction of the population of the host country (with possible exceptions of wars or other abrupt phase transitions, to minor degrees). Then, if one wants to study the social network of the host country, the conventional models cease to perform, because the fraction of new nodes to existing nodes does not tend to infinity, but is small. The same is true for any slowly-growing realistic network where the extrapolation of the near future provided information on the current state is called for.


\subsection{Organization of the Paper}
First in subsection~\ref{subsec:unif_single} we consider the uniform attachment scheme, with each newly introduced node linking to one existing node picked uniformly at random. We compare our results with the ones present in the literature. Then in subsection~\ref{subsec:unif_multiple} we consider the uniform attachment for multiple linking, where each new node connects to $\beta$ existing nodes drawn uniformly at random. 
In~\ref{sec:preferential} we examine the preferential attachment scheme. First in~\ref{subsec:pref_single} we consider each new node linking to only one existing node with probabilities assigned to existing nodes proportional to their degrees. Then in~\ref{subsec:pref_multiple} we assume each new node attaches to $\beta$ existing nodes. So each new node has degree $\beta$ upon birth. We solve for the expected degree distribution in all cases. Throughout the paper, we compare our theoretical findings with simulations.

\section{Uniform Attachment} \label{sec:uniform}

\subsection{Single Connection} \label{subsec:unif_single}

We start from an initial   graph at time $t=0$ with $N(0)$ nodes. We denote the degree distribution at the outset by $n_k$. At each timestep, a new node is introduced. It picks one of the existing nodes uniformly at random and connects to it. Nodes are added one by one. If the initial condition is a single node, the resulting graph will be the conventional Random Recursive Tree~\cite{RRTmath1, RRTmath2, rednerM1}.

 Let $ \alpha$ represent the rate at which new nodes are introduced, that is, $\alpha \Delta t $ nodes are added in a time interval of duration $\Delta t$. It also means that each node is added within~$\frac{1}{\alpha}$ unit times. So for example, if~${\alpha= 100 }$, then 100 nodes are introduced per unit time,  and each of them arrives at 0.01  unit times.   At time~$t$ there  are $N(t)=N(0)+\alpha t$ nodes. Let $N_k(t)$ denote the expected number of nodes whose degree is $k$ at time $t$. Let us focus on the expected variation in $N_k(t)$ in  the  time increment $\frac{1}{\alpha}$ within which one new node is  added.

With probability $\frac{ N_k(t)}{N(t)} $, a node with degree $k$ receives a link, and its degree increments. Consequently,  $N_k$ decrements and $N_{k+1}$ increments, both by one. Similarly, 
with probability $\frac{N_{k-1}(t)}{N(t)} $, a node with degree $k-1$ receives a link, hence  $N_k$ increments and $N_{k-1}$ decrements, both by one. So we have: 
\eq{
N_k (t+\frac{1}{\alpha})-N_k(t)=   \frac{N_{k-1}(t)-N_k(t)}{N(0)+\alpha t} 
.
}
Note that   the case of $k=1$ is distinct. Each new node increments $N_1$ by one. So, 
\eq{
N_1(t+\frac{1}{\alpha})-N_1(t)=-   \frac{N_1(t)}{N(0)+\alpha t} + 1 
.}
These two equations can be condensed into one: 
\begin{align}
N_k (t+\frac{1}{\alpha})-N_k(t) = \frac{ 1 }{N(0)+\alpha t} (N_{k-1}-N_k) 
+   \delta_{k,1}
,
\label{DELTA_Nk}
\end{align}
where $\delta_{k,1}$ is the Kronecker delta function (i.e., $\delta_{k,1} = 1$ if $k=1$, and $\delta_{k,1} = 0$ otherwise). 
Dividing both sides by $\frac{1}{\alpha}$, and denoting~$\frac{1}{\alpha}$ by~${\Delta t}$ (which means that one node arrives per $\Delta t$),  we can recast this equation as the following: 
\begin{align}
\frac{N_k (t+\Delta t )-N_k(t)}{\Delta t} = \frac{ \alpha }{N(0)+\alpha t} (N_{k-1}-N_k) 
+   \alpha \delta_{k,1}
.
\label{DELTA_Nk_2}
\end{align}
In the limit~${\Delta t \rightarrow 0}$,  the following differential equation is obtained for the dynamics of the expected degree distribution:
\eqq{
\dot{N_k}=\frac{\alpha }{N(0)+\alpha t} (N_{k-1}-N_k)+\alpha \delta_{k,1},
}{diffeq}
where $\dot{N_k}$ is the first derivative of $N_k(t)$ with respect to time, and explicit dependence on time is omitted for expositional simplification. 

Approximating the difference equation~\eqref{DELTA_Nk} with its differential analog~\eqref{diffeq} has error of order~$\Delta t$ (readily seen through the Taylor expansion of~${N_k(t+\Delta t}$)), which can be controlled by rescaling of time. Error shrinks as~$\alpha$ grows:
 \eq{
\dot{N_k}=\frac{\alpha }{N(0)+\alpha t} (N_{k-1}-N_k)+\alpha \delta_{k,1}+ O\left( \displaystyle \frac{1}{\alpha} \right).
}
This continuous approximation is justified more rigorously using martingales in~\cite{martin_1,martin_2} (also see \cite{martin_3,martin_4}). In this paper the goodness of this approximation is empirically verified through simulations. 


Note that, from (7), we see that the increments $N_k(t + \frac{1}{\alpha}) - N_k(t)$ take rational values proportional to $\frac{1}{N(0) + \alpha t}$. Approximating the left-hand side with a differential yields (9). Since the denominator of (7) has the factor $N(0) + \alpha t$, the approximation becomes more accurate as $t$ grows, hence increasing $N(0) + \alpha t$. The approximation is also more accurate when $N(0)$ is large. When both $N(0)$ and $t$ are small, then the continuous approximation in the $N_k$ domain becomes less accurate. Note that, as long as $N(0)$ is large, $t$ need not be large. This is particularly important for applications where networks are already large (such as those mentioned in Section~I), and one would like to predict the short-term evolution of the degree distribution. In these settings, the expressions obtained throughout this paper are applicable for any time regime.


To solve~\eqref{diffeq}, we use the generating function $\psi(z,t)=\sum_k z^{-k} N_k(t)$, which is the conventional Z-transform in the $k$ domain. Using~\eqref{diffeq} we get
\eqq{
\frac{\partial \psi(z,t)}{\partial t}=\frac{\alpha }{N(0)+\alpha t}(z^{-1}-1)\psi(z,t)+\frac{\alpha}{z}.
}{psi_Z_unif}
So we intend to solve the following differential equation in the time domain: 
\eqq{
\frac{\partial \psi(z,t)}{\partial t}-\frac{\alpha (z^{-1}-1)}{N(0)+\alpha t}\psi(z,t)= \frac{\alpha}{z}.
}{psi_PDE_1}
After solving this equation and applying the initial conditions 
(Appendix~\ref{app: solve_PDE_unif_single}), we obtain the generating function:

\begin{align}
\psi(z,t)&=\frac{N(0)+\alpha t}{z(2-\frac{1}{z})}+\frac{N(0)}{N(0)+\alpha t} \left[\psi(z,0) (1+\frac{\alpha t}{N(0)})^{\frac{1}{z}}\right] \nonumber \\
&-
\frac{N(0)^2}{N(0)+\alpha t} \frac{1}{z(2-\frac{1}{z})}\left(1+\frac{\alpha t}{N(0)}\right)^{\frac{1}{z}}.
\label{psi_unif_Z_text}
\end{align}
To take the inverse transform, first note that: 
\eq{
\sum_{k=1}^{\infty} \frac{1}{2^k} z^{-k}=\frac{1}{z(2-\frac{1}{z})}
.}
Also, denoting $\frac{\alpha t}{N(0)}$ by $\lambda$, note that we have: 
\eq{
(1+\lambda)^{\frac{1}{z}}= e^{\frac{\ln(1+\lambda)}{z}} 
.}
Using the Taylor expansion of the exponential, we get: 
\eq{
(1+\lambda)^{\frac{1}{z}}= \sum_{k=0}^{\infty}  \frac{\big[\ln(1+\lambda)\big]^k}{k!} z^{-k}
.}
So the inverse transforms are:
\eq{
\begin{cases}
\displaystyle \frac{1}{z(2-\frac{1}{z})} \xrightarrow{\mathcal{Z}^{-1}} \frac{1}{2^k} u(k-1) \\ \\
\displaystyle \left(1+\frac{\alpha t}{N(0)}\right)^{\frac{1}{z}} \xrightarrow{\mathcal{Z}^{-1}} 
\frac{\bigg[ \ln[1+\frac{\alpha t}{N(0)}] \bigg]^k}{k!},
\end{cases}
}
where $u(x)$ is the Heaviside step function (i.e., $u(x) = 0$ for $x < 0$, and $u(x) = 1$ for $x\ge0$). Finally, note that the multiplication of Z-transforms yields convolution after inversion. Let us denote the degree distribution of the initial graph by $n_k$, that is, $n_k$ is the fraction of nodes at the outset with degree $k$. So by inverting~\eqref{psi_unif_Z_text}, for~$ t \geq 0$ and~$1 \leq  k \leq  N(0)$ we obtain: 
\begin{align}
&N_k(t)=\frac{N(0)+\alpha t}{2^k} u(k-1) \nonumber \\
&+ \frac{N(0)^2}{N(0)+\alpha t} \left\{n_k * \frac{\left[ \ln \left( 1+\frac{\alpha t}{N(0)} \right) \right]^k}{k!} \right\} \nonumber \\
&-\frac{N(0)^2}{N(0)+\alpha t} \left\{ \left( \frac{u(k-1)}{2^k} \right) * \frac{\left[ \ln \left( 1+\frac{\alpha t}{N(0)} \right) \right]^k}{k!} \right\},  
\label{Nk_final_sec1}
\end{align}
where $*$ denotes the convolution operator. For general sequences $a_k$ and $b_k$, the convolution is another sequence in the $k$ domain which is  defined as follows: 
\al{
(a*b)_k \stackrel{\text{def}}{=} \displaystyle \sum_{\eta=-\infty}^{+\infty} a_{\eta} b_{k-\eta}.
}
If the sequences are zero for negative values of $k$ (which is the case in our problem),  this can be simplified to: 
\al{
(a*b)_k = \displaystyle \sum_{\eta=0}^{k} a_{\eta} b_{k-\eta}.
}

 To get the degree distribution, we divide the result  in~\eqref{Nk_final_sec1} by the total number of nodes at time $t$, which is equal to $N(0)+\alpha t$. So   for~$ t \geq 0$ and~$1 \leq  k \leq  N(0)$ we obtain:
\begin{align}
&p_k(t)=\frac{u(k-1)}{2^k} \nonumber \\
& + \left(\frac{N(0)}{N(0)+\alpha t}\right)^2 \left\{n_k * \frac{\left[ \ln \left( 1+\frac{\alpha t}{N(0)} \right) \right]^k}{k!} \right\} \nonumber \\
&-\left(\frac{N(0)}{N(0)+\alpha t}\right)^2 \left\{ \left( \frac{u(k-1)}{2^k} \right) * \frac{\left[ \ln \left( 1+\frac{\alpha t}{N(0)} \right) \right]^k}{k!} \right\}.
 \label{result1}
\end{align}
As $t \rightarrow \infty$, the effect of initial conditions vanish.  So the first term dominates. In this limit,   the asymptotic  degree distribution is:
\eq{
\lim_{t \rightarrow \infty} p_k(t) =\frac{1}{2^k}.
}
Note that this matches the asymptotic behavior previously found for RRTs as presented in \cite{RRTmath1, RRTmath2, rednerM1}.

\begin{figure}[t]
  \centering
  \includegraphics[width=\columnwidth]{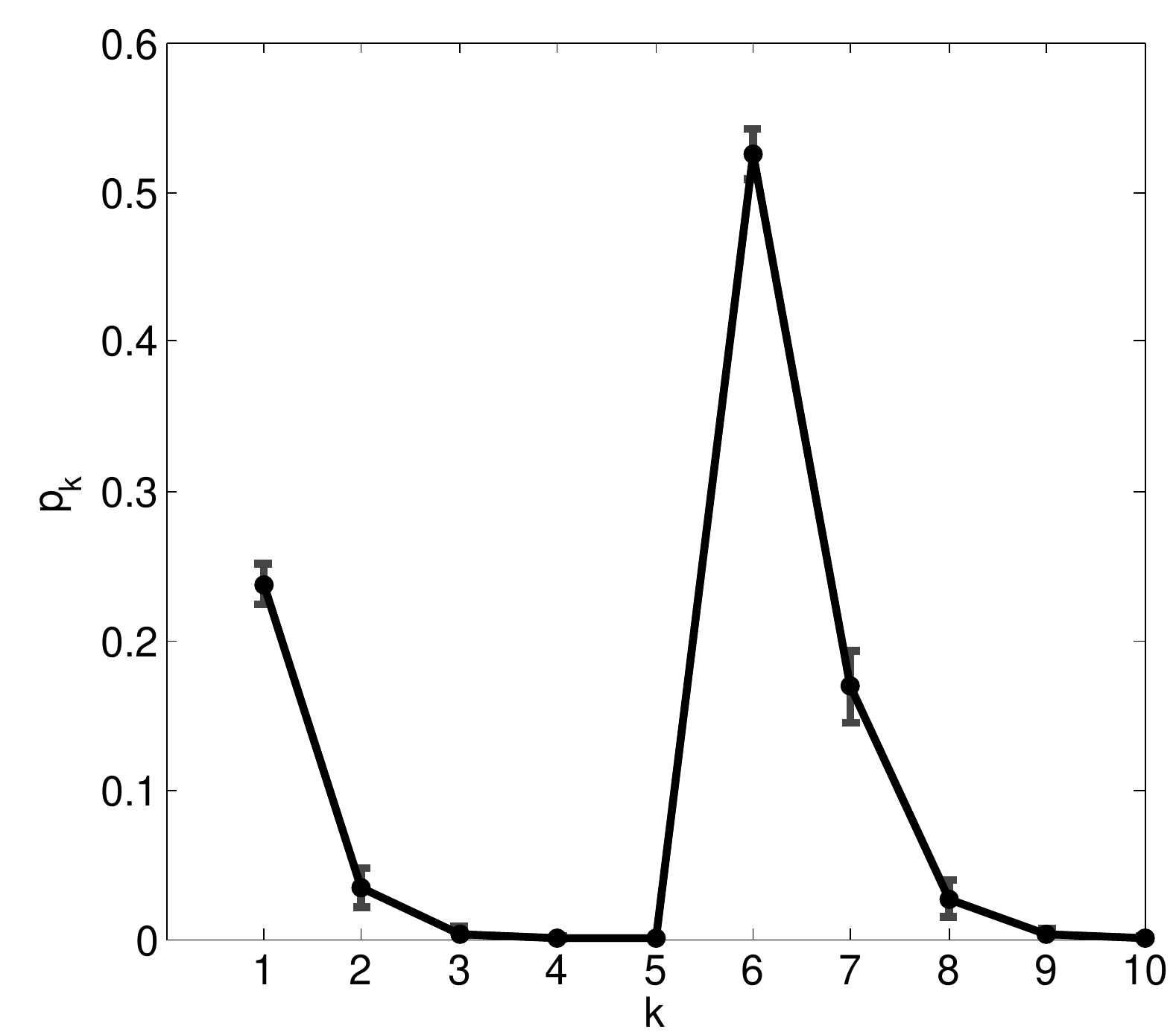}
  \caption[Figure 1]%
  {Degree distribution at time $t=20$. The initial graph is a 6-regular ring with ${N(0)=50}$ nodes. Growth is under uniform probabilities and single attachments. At each timestep, one node is added, so ${\alpha=1}$. The error bars for  50 Monte~Carlo trials are plotted, along with the theoretical curve. }
\label{fig_unif_single_deg}
\end{figure}

To compare theoretical prediction with simulation results, first we start off with a 
 6-regular  graph. This means that all nodes have 6 neighbors. We build this graph by first making a ring of 50 nodes, and then connect each node to the pairs of  second and third closest neighbors. 
Figure~\ref{fig_unif_single_deg}  shows the degree distribution at time $t=20$, that is, $p_k(t=20)$. The results are average over 50 Monte~Carlo trials. Also, $\alpha=1$, so nodes are introduced one at a time. As can be seen in the figure, the second majority belongs to  degree 1, which are the newly born nodes. Nodes of degree 6 are mostly the initial ones who have received no new link yet, and those with degree 7 have  received only one additional link.

\begin{figure}[t]
  \centering
  \includegraphics[width=1\columnwidth]{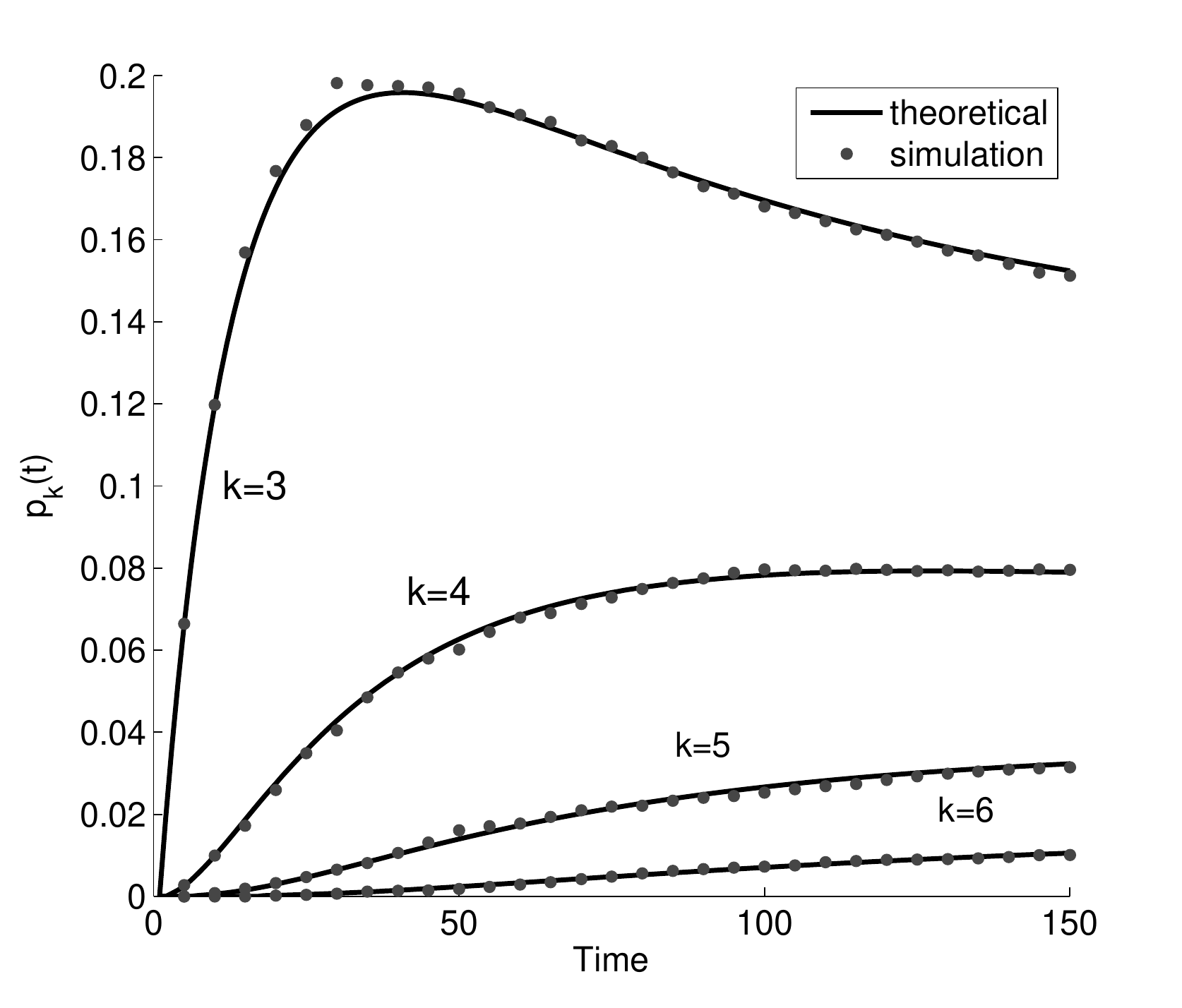}
  \caption[Figure 1]%
  {Growth under uniform probabilities and single attachments, on a ring (2-regular) of 50 nodes . The fractions of nodes having degree 3,4,5,6  are depicted in time. ${\alpha=1}$ is used. The results are average over 20 Monte~Carlo trials. }
\label{fig_unif_single}
\end{figure}

\begin{figure}[t]
  \centering
  \includegraphics[width=1\columnwidth ]{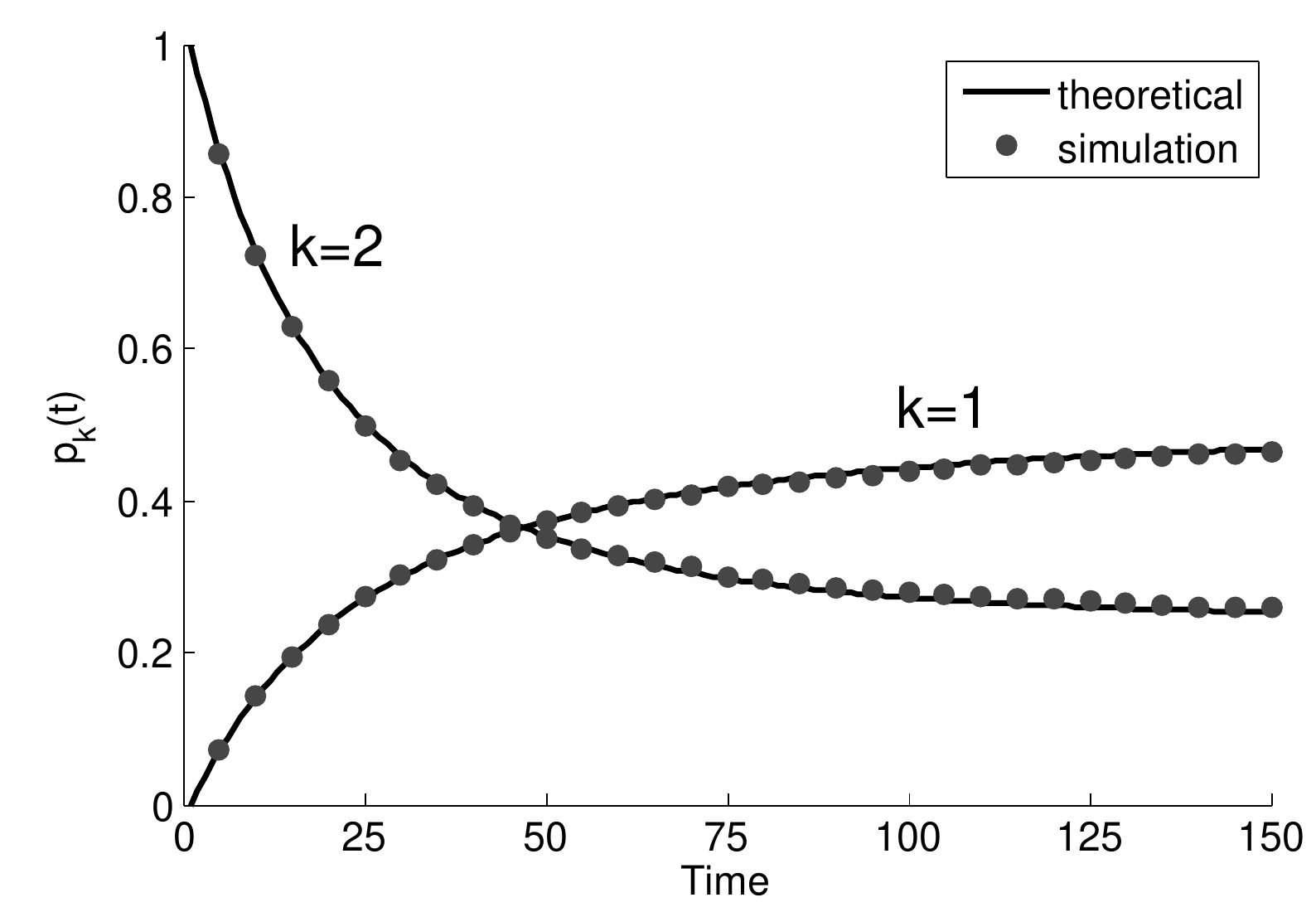}
  \caption[Figure 1]%
  {Growth under uniform probabilities and single attachments, on a ring (2-regular) of 50 nodes . The fractions of nodes having degree 1 and 2  are depicted in time. ${\alpha=1}$ is used. The results are average over 20 Monte~Carlo trials. }
\label{fig_k_1_2}
\end{figure}

For the next simulation, we take a  ring (2-regular) of 50 nodes and  we plot $p_k(t)$ as a function of time, for $k=3,4,5,6$. The simulations and theoretical results are shown in Figure~\ref{fig_unif_single}.  Nodes with degree 3 are the ones who have received one link from the newly added nodes, who outnumber  those who have received two, as seen in the figure. This is expected because initially it is less probable that the new node attaches to a node who already have received a link  than a node whose degree is still~$2$ , since the latter out-numbers the former at early times, thus its population  grows substantially. After a while, many nodes have degree 3 and now that they receive new links, their degrees turn 4, reducing the population of degree-3 nodes. Figure~\ref{fig_k_1_2} shows $p_k(t)$ for the case of $k=1,2$. It can be seen that the fraction of nodes with degree $2$ decreases and that of those with degree 1 increases. The reason is tat each new node that is added to the network has degree 1, and those existing nodes with degree 2 receive links fro the new nodes and their degrees increment, and is not 2 anymore, hence the decline in the population of nodes with degree 2. 

Next, the mean and variance of simulations are tabulated to provide estimates for fluctuations around the mean values which are solved for. Table~\ref{table_unif_single_attachment} presents these values for a 4-regular ring of 20 nodes, at different times.

\begin{table}[t]
\centering 
\begin{tabular}{| c |c |c |c|c|} 
\hline 
 \backslashbox{Time}{Degree} & $k=3$ & $k=4$ & $k=5$  & $k=6$\\ [0.5ex] 
\hline
t=5  & \dcell{0.0004\\ 0.0000} & \dcell{0.6901\\ 0.0003}& \dcell{0.1327\\ 0.0008}& \dcell{0.0090\\ 0.0002}\\     \hline
t=10  & \dcell{0.0042\\ 0.0001} & \dcell{0.4702\\ 0.0008}  & \dcell{0.1831\\ 0.0016}& \dcell{0.0320\\ 0.0004} \\ \hline
t=15  & \dcell{0.0103\\ 0.0001} & \dcell{0.3415\\ 0.0009} & \dcell{0.1893\\ 0.0016}& \dcell{0.0493\\ 0.0005}   \\ \hline
t=20  & \dcell{0.0177\\ 0.0002} & \dcell{0.2604\\ 0.0008} & \dcell{0.1801\\ 0.0014}& \dcell{0.0602\\ 0.0005}    \\ \hline
t=25  & \dcell{0.0251\\ 0.0002} & \dcell{0.2062\\ 0.0008} & \dcell{0.1669\\ 0.0012}& \dcell{0.0661\\ 0.0005}     \\ \hline
t=30  & \dcell{0.0321\\ 0.0002} & \dcell{0.1689 \\ 0.0007} & \dcell{0.1525\\ 0.0010}& \dcell{0.0688\\ 0.0005}   \\ \hline
t=35  & \dcell{0.0385\\ 0.0003} & \dcell{0.1421\\ 0.0006}& \dcell{0.1392\\ 0.0008}& \dcell{0.0697\\ 0.0004}    \\ \hline
t=40  & \dcell{0.0447\\ 0.0003} & \dcell{0.1224\\ 0.0006} & \dcell{0.1272\\ 0.0007}& \dcell{0.0693\\ 0.0004}   \\
 [1ex] 
\hline 
\end{tabular}
\caption{Uniform single attachment: the mean and variance (top and bottom row of each cell, respectively) of $p_k(t)$ for different instants of time and $k=3,4,5,6$. The substrate is a 4-regular ring of 20 nodes. the ensemble consists of 4000 Monte~Carlo trials.} 
\label{table_unif_single_attachment} 
\end{table}

%

\subsection{Multiple Connections} \label{subsec:unif_multiple}

Now, let us consider multiple attachments. Each new node that is introduced, chooses $\beta$ existing nodes (where~${\beta \geq 1}$ is an integer) uniformly at random and links to them. An essential difference of this scheme from the previous one is that, if one starts from a disconnected graph, then the probability of ending up with a connected graph is nonzero. This probability was zero in the previous case, because each newly-introduced node only linked to one existing node and could not make a connection between two disconnected components. Also note that in this case one must have~${\beta \leq N(0)}$, so that the growth mechanism can start off. Otherwise, link multiplicity arises, that is, more than a link  should be allowed between two nodes, which is tacitly assumed not to be the case throughout. 

Taking the similar steps that led to~\eqref{DELTA_Nk}, the change in $N_k(t)$ is given by: 
\eq{
\frac{N_k (t+\Delta t)-N_k(t) }{ \Delta t}= \frac{\beta \alpha }{N(0)+\alpha t} (N_{k-1}-N_k) 
+ \alpha \delta_{k,\beta}
.}

Note that the last term indicates that each new node adds one to $N_\beta$, because its degree is~$\beta$.  The differential equation analog for $N_k(t)$ becomes
\eq{
\dot{N_k}=\frac{\beta \alpha }{N(0)+\alpha t} (N_{k-1}-N_k)+\alpha \delta_{k,\beta} ~.
}
Taking the Z-transform, we get
\eq{
\frac{\partial \psi(z,t)}{\partial t}=\frac{\alpha \beta}{N(0)+\alpha t}(z^{-1}-1)\psi(z,t)
+\frac{\alpha}{z^\beta}.
}
So we arrive at the following differential equation:  
\eqq{
\frac{\partial \psi(z,t)}{\partial t}-\frac{\beta \alpha (z^{-1}-1)}{N(0)+\alpha t}\psi(z,t)= \frac{\alpha}{z^\beta}.
}{psi_PDE_2}
The solution procedure is given in~\ref{app: solve_PDE_unif_multiple}. The generating function is
\begin{align}
\psi(z,t)&=\frac{1}{z^\beta} \frac{N(0)+\alpha t}{1+\beta(1-z^{-1})} \nonumber \\
&+ \psi(z,0) \bigg[\frac{N(0)}{N(0)+\alpha t} \bigg]^{\beta(1-z^{-1})} \nonumber \\
&- \frac{N(0)}{z^\beta \big[ 1+\beta(1-z^{-1})\big] }  \bigg[\frac{N(0)}{N(0)+\alpha t} \bigg]^{\beta(1-z^{-1})}
.
\label{psi_unif_Z_2}
\end{align}
Now we must invert this, term by term. This is done in appendix~\ref{app: Z_unif_multiple}. After inversion, 
  for~$ t \geq 0$ and~$1 \leq  k \leq  N(0)$ we obtain: 
\begin{align}
&N_k(t)=
\frac{N(0)+\alpha t}{\beta} \left( \frac{\beta}{\beta+1} \right)^{k-\beta+1} u(k-\beta)
\nonumber \\
&+ \left[ \frac{N(0)^{\beta+1}}{(N(0)+\alpha t)^{\beta}} \right]    \left\{ n_k * \frac{\left[\beta  \ln \left( 1+\frac{\alpha t}{N(0)} \right) \right]^k}{k!} \right \}
\nonumber \\
&-\left[ \frac{N(0)^{\beta+1}}{\beta(N(0)+\alpha t)^{\beta}} \right] \times \nonumber \\
&\left\{ \left( \frac{\beta}{\beta+1} \right)^{k-\beta+1} u(k-\beta)  * \frac{\left[\beta  \ln \left( 1+\frac{\alpha t}{N(0)} \right) \right]^k}{k!} \right\}.
 \end{align}
Then  we divide by the total number of nodes~${N(t)=N(0)+\alpha t}$ to get the degree distribution  for~$ t \geq 0$ and~$1 \leq  k \leq  N(0)$. The result is:
\begin{align}
&p_k(t)=
\frac{1}{\beta} \left( \frac{\beta}{\beta+1} \right)^{k-\beta+1} u(k-\beta) \nonumber \\
&+ \left( \frac{N(0)}{N(0)+\alpha t} \right)  ^{\beta+1}     \left\{ n_k * \frac{\left[\beta  \ln \left( 1+\frac{\alpha t}{N(0)} \right) \right]^k}{k!} \right \}
\nonumber \\
&-\left( \frac{N(0)}{N(0)+\alpha t} \right)  ^{\beta+1}  \frac{1}{\beta} 
\times \nonumber \\
& \left\{ \left( \frac{\beta}{\beta+1} \right)^{k-\beta+1} u(k-\beta)  * \frac{\left[\beta  \ln \left( 1+\frac{\alpha t}{N(0)} \right) \right]^k}{k!} \right\}, 
\end{align}

Now let us look at the long-time behavior of the result. When ${t \rightarrow \infty}$, the second and the third terms vanish. The first term prevails and tends to the following:
\begin{equation}
p_k(t) \sim \frac{1}{\beta} \left( \frac{\beta}{\beta+1} \right)^{k-\beta+1} u(k-\beta).
\end{equation}
Note that for the case of $\beta = 1$, the same asymptotic distribution is obtained as the previous section. Also note that in the asymptotic limit, all nodes have degree at~least~$\beta$ and the fraction of nodes with degree less than~$\beta$  tends to zero. 

\begin{figure}[t]
  \centering
  \includegraphics[width=1\columnwidth]{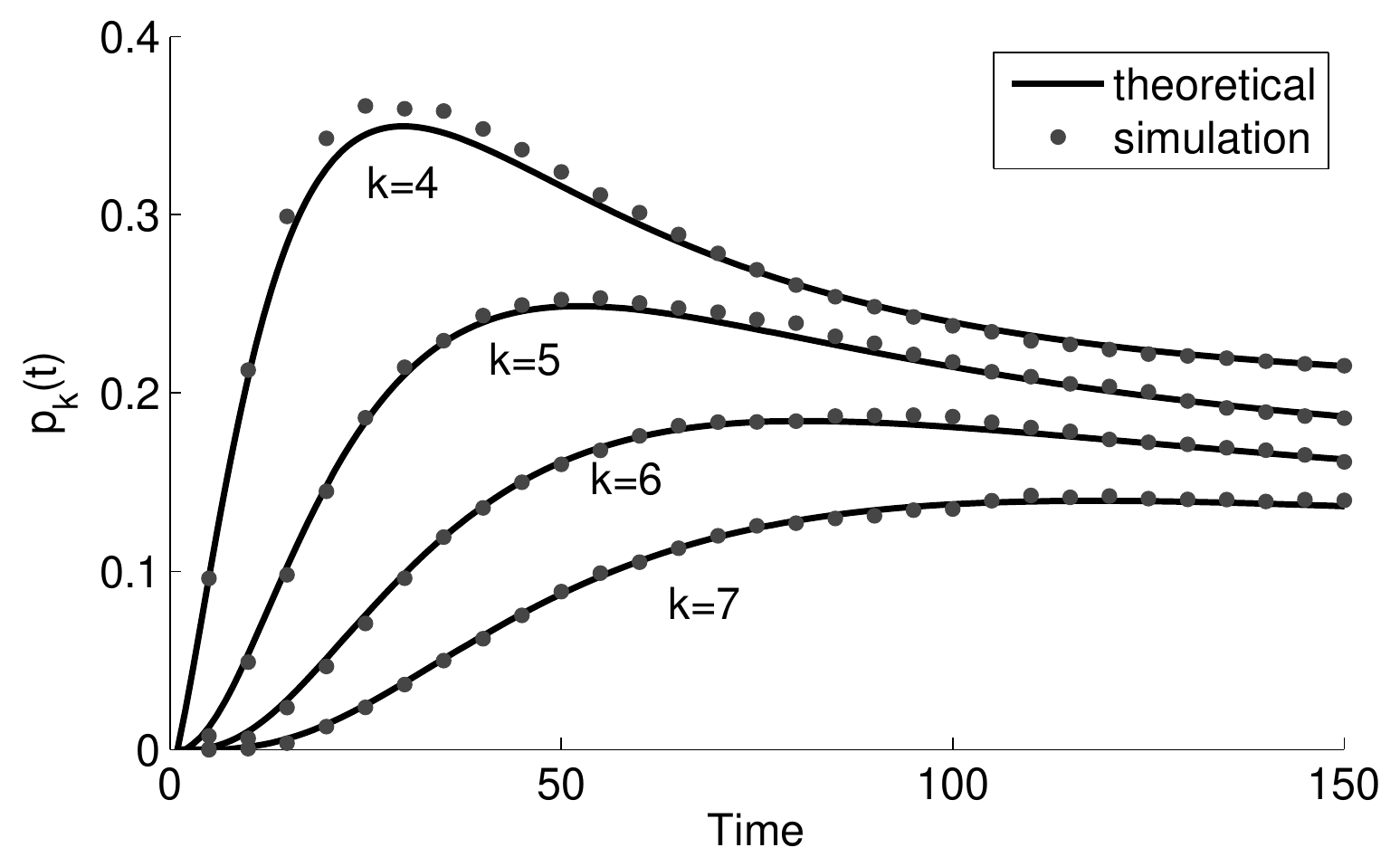}
  \caption[Figure 1]%
  {Growth under uniform probabilities and multiple attachments with $\beta=3$ and ${\alpha=1}$, on a ring (2-regular) of 30 nodes.  The fractions of nodes having degree 4,5,6,7  are depicted in time. The results of 30 Monte~Carlo trials are plotted, along with the theoretical curve.}
\label{fig_unif_multiple}
\end{figure}

Figure~\ref{fig_unif_multiple} shows the simulation results for  a ring~(2-regular)   of 30 nodes, and $p_k(t)$ is depicted versus time, for $k=4,5,6,7$.  The value of ${\beta}$ is 3. The number of Monte~Carlo trials is 30.  It can be seen in the graph that the nodes with degree 4, who are mostly the initial nodes who have received one link from the newcomers,  outgrow those with degree 5. This is because they also outnumber them, giving them greater link reception probabilities. After a while, this trend declines because many nodes will have degree 4, and now that they receive a new link, they will turn into  nodes of degree 5, enhancing  the growth of the degree 5 nodes, diminishing the portion of nodes with degree 4. Similarly, the overshoot of degree 6 curve happens after that of degree 5, and so on.

\begin{figure}[t]
  \centering
  \includegraphics[width=1\columnwidth]{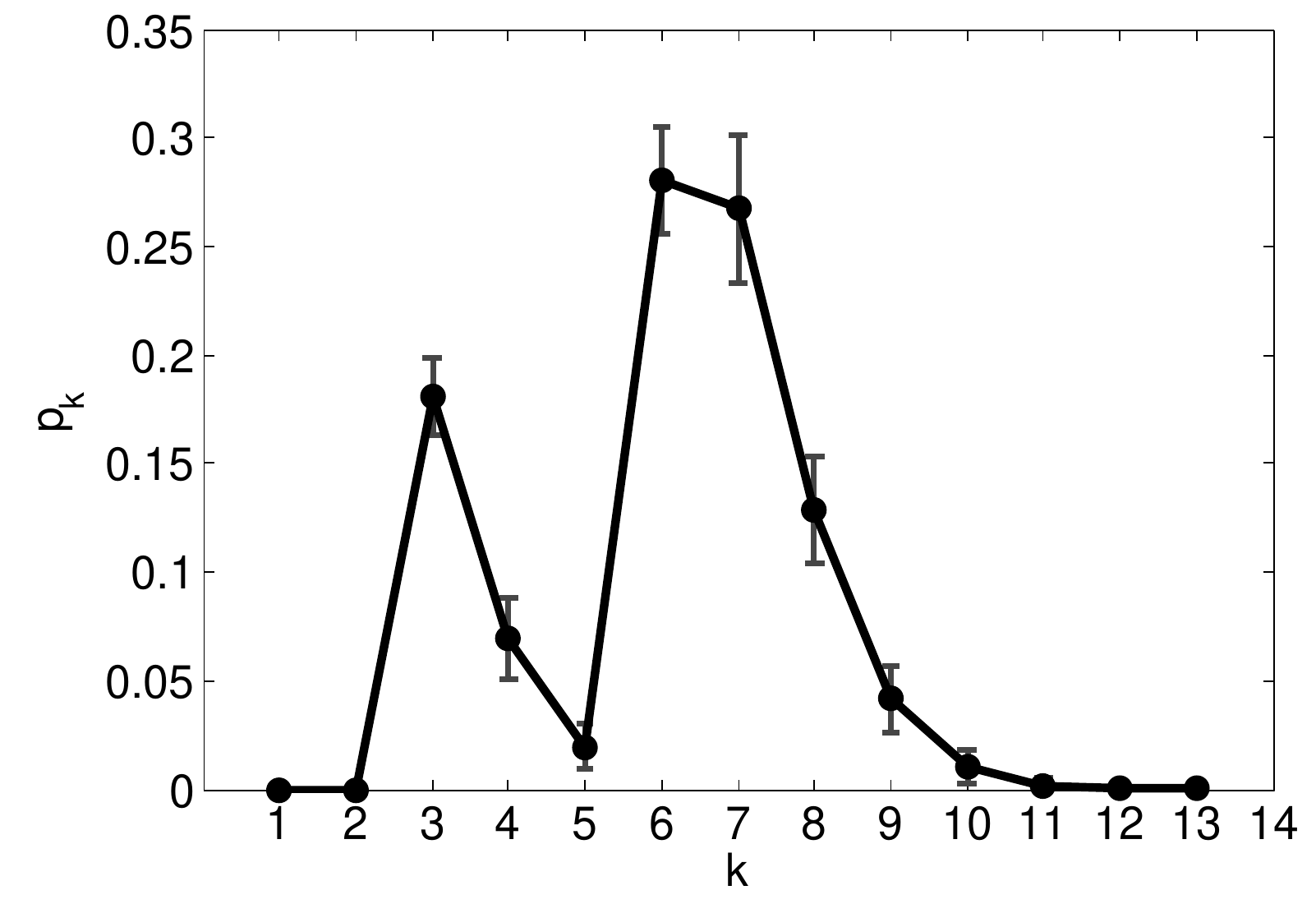}
  \caption[Figure 1]%
  {Degree distribution at time $t=20$ for a 6-regular ring of 50 nodes, subject to growth with uniform probabilities and multiple attachments. The value of $\beta$ is 3 and ${\alpha=1}$. The error bars for 50 Monte~Carlo trials are plotted along with the theoretical curve.}
\label{fig_unif_multiple_deg}
\end{figure}

Figure~\ref{fig_unif_multiple_deg}  shows $p_k(t)$ for $t=20$, for a  6-regular ring of total 50 nodes. The number of Monte~Carlo trials is 50. The value of ${\beta}$ is 3. The peak at ~${k=3}$ seen in the figure  is due to the newly added nodes, who all have degree 3. Most of the initial nodes have received zero or one links by this time, hence the other  peak at~$k=6,7$.

The mean and variance of simulations are presented  in Table~\ref{table_unif_mult}.

\begin{table}[t]
\centering 
\begin{tabular}{| c |c |c |c|c|} 
\hline 
 \backslashbox{Time}{Degree} & $k=3$ & $k=4$ & $k=5$  & $k=6$\\ [0.5ex] 
\hline
t=5  & \dcell{0.1343\\ 0.0004} & \dcell{0.4921\\ 0.0014}& \dcell{0.2907\\ 0.0022}& \dcell{0.0718\\ 0.0006}\\     \hline
t=10  & \dcell{0.1966\\ 0.0005} & \dcell{0.2939\\ 0.0016}  & \dcell{0.2863\\ 0.0020}& \dcell{0.1562\\ 0.0012} \\ \hline
t=15  & \dcell{0.2216\\ 0.0006} & \dcell{0.2289\\ 0.0014} & \dcell{0.2412\\ 0.0016}& \dcell{0.1774\\ 0.0012}   \\ \hline
t=20  & \dcell{0.2335\\ 0.0006} & \dcell{0.2043\\ 0.0012} & \dcell{0.2062\\ 0.0012}& \dcell{0.1745\\ 0.0012}    \\ \hline
t=25  & \dcell{0.2400\\ 0.0005} & \dcell{0.1939\\ 0.0010} & \dcell{0.1835\\ 0.0010}& \dcell{0.1639\\ 0.0010}     \\ \hline
t=30  & \dcell{0.2430\\ 0.0005} & \dcell{0.1905 \\ 0.0008} & \dcell{0.1692\\ 0.0008}& \dcell{0.1522\\ 0.0008}   \\ \hline
t=35  & \dcell{0.2452\\ 0.0004} & \dcell{0.1883\\ 0.0008}& \dcell{0.1600\\ 0.0008}& \dcell{0.1432\\ 0.0007}    \\ \hline
t=40  & \dcell{0.2464\\ 0.0004} & \dcell{0.1880\\ 0.0006} & \dcell{0.1536\\ 0.0007}& \dcell{0.1358\\ 0.0006}   \\
 [1ex] 
\hline 
\end{tabular}
\caption{Uniform multiple attachment: the mean and variance (top and bottom row of each cell, respectively) of $p_k(t)$ for different instants of time and $k=3,4,5,6$. The substrate is a 4-regular ring of 20 nodes. The value of~$\beta$ is 3. the ensemble consists of 4000 Monte~Carlo trials.} 
\label{table_unif_mult} 
\end{table}


\section{Preferential Linking} \label{sec:preferential}
In this section we are going to focus on preferential attachment. New nodes, instead of selecting from the existing nodes uniformly at random, assign to them probabilities of connection,  proportional to their degrees. So, each existing node has the chance of receiving a link from the newly-introduced node equal to its degree, divided by the sum of the degrees of every existing node. 
First we will assume the case where a new node only attaches to a single existing node, and then the case of multiple connections is considered.

\subsection{Single Connection} \label{subsec:pref_single}

As mentioned above, in the preferential attachment scheme, an existing node with degree $k$ receives a link with probability $k/\sum_{\ell} \ell N_{\ell}$, where the denominator is the sum of the degrees of all existing nodes. So the probability that the destination node selected by a newly-born node has degree $k$  is equal to $kN_k/\sum_{\ell} \ell N_{\ell}$.  Using the same approach that led to~\eqref{diffeq}, we arrive at the following differential equation for the  evolution of $N_k(t)$: 
\eqq{
\dot{N_k}=\frac{\alpha }{\sum_{\ell} \ell N_{\ell}} ((k-1)N_{k-1}-k N_k)+\alpha \delta_{k,1}.
}{Nkdot2}
Now to proceed as before, we take the Z-transform of this equation. First note that if $X(z)$ is the Z-transform of a discrete function $x_k$, then $-z\frac{dX(z)}{dz}$ is the Z-transform for the the function $kx_k$. This means that, if the Z-transform of $N_k$ is $\psi(z)$, then the Z-transform of the first term on the right hand side is as follows:
\eq{
(k-1)N_{k-1}-k N_k \xrightarrow{\mathcal{Z}} (z-1) \frac{d\psi(z)}{dz}
.}
Second, note that the denominator of the attachment probabilities, 
$\sum_{\ell} \ell N_{\ell}$, is twice the number of links in the graph. Let us denote the number of links in the graph by $L(t)$. Note that, since each new node adds one new link,  we have: 
\eq{
L(t)=L(0)+\alpha t \Longrightarrow 2L(t)=2L(0)+2\alpha t
.}

Twice the number of links in the initial graph equals $N(0)\bar{k}_0$, 
where $\bar{k}_0$ denotes the average degree of the initial graph. So we get: 
\eq{
2L(t)=N(0)\bar{k}_0+2\alpha t \Longrightarrow \sum_{\ell} \ell N_{\ell}=N(0)\bar{k}_0+2\alpha t
.}
We will temporarily use 
\eqq{
\lambda \stackrel{\text{def}}{=} N(0)\bar{k}_0
}{lambda_def}
for brevity. The Z-transform of~\eqref{Nkdot2} is:
\eqq{
\frac{\partial \psi}{\partial t}- \frac{\alpha (z-1)}{\lambda + 2\alpha t} 
\frac{\partial \psi}{\partial z}= \alpha z^{-1}
.}{diff_eq_pref}
This is a first-order partial differential equation. We  solve this equation using the method of characteristics. For the convenience of the reader, we briefly shed light on how this method works through a simple example in appendix~\ref{app:PDE} (we refer the reader to~\cite{z1,z2,z3}, or other elementary references on partial differential equations, for further details), and then provide the solution in appendix~\ref{app: PDE_pref_single}, where we obtain:
\begin{align}
\displaystyle 
&\psi(z,t)= \psi_0\bigg[(z-1)\sqrt{\frac{\lambda+2\alpha t}{\lambda}}+1\bigg] - 2\alpha t(z-1)+\alpha t \nonumber \\ 
&-2\alpha 2(z-1)^2\ln(1-z^{-1}) +(z-1) \lambda \sqrt{\frac{\lambda+2\alpha t}{\lambda}}\nonumber \\
& - \lambda(z-1)-\lambda(z-1)^2\ln(1-z^{-1})
\nonumber \\
&-(z-1)^2(\lambda+2\alpha t) \ln \bigg[ 1+\frac{1}{z-1}\sqrt{\frac{\lambda}{\lambda+2\alpha t}}\bigg]
.
\label{psi_temp_2}
\end{align}
Note that from the argument of the logarithm, we know the region of convergence of the Z-transform is $z>1$, since the logarithm is not defined otherwise (this agrees with what one would expect intuitively, that since~$N_k(t)$ is zero for~$k<0$ by definition, the region of convergence would be~$z>1$). Now let us define the new variable
\eqq{
c(t) \stackrel{\text{def}}{=} 1-\sqrt{\frac{\lambda}{\lambda+2\alpha t}}
.}{c_def}
This quantity is positive and less than unity at all times. Now, note that we have: 
\eq{ 
\begin{cases}
\displaystyle 1+\frac{1}{z-1}\sqrt{\frac{\lambda}{\lambda+2\alpha t}} = \frac{1-cz^{-1}}{1-z^{-1}} \\ \\
\displaystyle (z-1)\sqrt{\frac{\lambda+2\alpha t}{\lambda}}+1= \frac{z-c}{1-c}
\end{cases}
.}
So we simplify~\eqref{psi_temp_2} further and arrive at: 
\begin{align}
\displaystyle 
&\psi(z,t)= \psi_0 \left( \frac{z-c}{1-c} \right)
 - 2\alpha t(z-1)+\alpha t \nonumber \\ 
&-(\lambda+2\alpha t) (z-1)^2 \ln(1-z^{-1}) + \lambda (z-1)\sqrt{\frac{\lambda+2\alpha t}{\lambda}} \nonumber \\
&-\lambda(z-1)-(\lambda+2\alpha t) (z-1)^2 \ln\left( \frac{1-cz^{-1}}{1-z^{-1}} \right).
\label{psi_temp_3}
\end{align}
Note that the two ${\ln(1-z^{-1})}$ terms cancel out. Also, note that there are three terms having the factor ${(z-1)}$. These three terms add up to: 
\begin{align}
\displaystyle 
&- 2\alpha t(z-1)+ \lambda (z-1)\sqrt{\frac{\lambda+2\alpha t}{\lambda}} -\lambda(z-1)
\nonumber \\
&= (z-1) \sqrt{\lambda+2\alpha t} \bigg( \sqrt{\lambda}- \sqrt{\lambda+2\alpha t} \bigg) \nonumber \\
&= -(z-1)(\lambda+2\alpha t) c.
\end{align}
These simplifications transform~\eqref{psi_temp_3} into the following: 
\begin{align}
\displaystyle 
\psi(z,t)&= \psi_0 \left( \frac{z-c}{1-c} \right)
 +\alpha t   -(z-1)(\lambda+2\alpha t) c \nonumber \\
&-(\lambda+2\alpha t) (z-1)^2 \ln\left(1-cz^{-1} \right).
\label{psi_temp_4}
\end{align}
We find the inverse Z-transform of this expression in appendix~\ref{app:Z_pref_single}. The result is:  
\begin{align}
\displaystyle 
N_k(t)&= \sum_{\ell}  N_{\ell}(0) (1-c)^{\ell} c^{k-\ell} \binom{k-1}{\ell-1} \nonumber \\
&+ (\lambda+2\alpha t) \bigg( \frac{c^k}{k}-2\frac{c^{k+1}}{k+1}+\frac{c^{k+2}}{k+2}\bigg) .
\label{psi_temp_7_1}
\end{align}
Replacing $\lambda$ by $N(0) \bar{k}_0$, for~${t \geq 0}$ and ${1 \leq  k \leq  N(0)}$we get:
 \begin{align}
\displaystyle 
N_k(t)&= \sum_{\ell} N_{\ell}(0) \binom{k-1}{\ell-1}   (1-c)^{\ell} c^{k-\ell}  \nonumber \\
&+ (N(0) \bar{k}_0+2\alpha t) \bigg( \frac{c^k}{k}-2\frac{c^{k+1}}{k+1}+\frac{c^{k+2}}{k+2}\bigg),  \nonumber \\
&   t \geq 0,  1 \leq  k \leq  N(0).
\label{psi_temp_7}
\end{align}
Now to get the degree distribution,  we divide this expression by the number of nodes at time $t$, which is equal to ${N(0)+\alpha t}$. As above, we denote the degree distribution of the initial graph by $n_k$, that is, $n_k$ is the fraction of nodes at the outset with degree $k$.  Thus the  final result for the degree distribution  for~${t \geq 0}$ and ${1 \leq  k \leq  N(0)}$ is the following: 
  \begin{align}
\displaystyle 
p_k(t)&= \frac{N(0)}{N(0)+\alpha t}\sum_{\ell}  n_{\ell} \binom{k-1}{\ell-1} (1-c)^{\ell} c^{k-\ell}  \nonumber \\
&+ \frac{N(0) \bar{k}_0+2\alpha t}{N(0)+\alpha t} \bigg( \frac{c^k}{k}-2\frac{c^{k+1}}{k+1}+\frac{c^{k+2}}{k+2}\bigg) ,  \nonumber \\
.
\label{pk_pref_single}
\end{align}
Now to find the asymptotic limit of this expression, first by combining~\eqref{c_def} and~\eqref{lambda_def}, we get
\eq{
c= 1-\sqrt{\frac{N(0) \bar{k}_0}{N(0) \bar{k}_0+2\alpha t}}.
}
Now note that as $t \rightarrow \infty$  we have: 
\eq{
\begin{cases}
\displaystyle \lim_{t \rightarrow \infty}  \frac{N(0)}{N(0)+\alpha t}=0 \\ \\
\displaystyle \lim_{t \rightarrow \infty} \frac{N(0) \bar{k}_0+2\alpha t}{N(0)+\alpha t} =2 \\ \\
\displaystyle \lim_{t \rightarrow \infty} c = 1.
\end{cases}
}
 Thus, the asymptotic behavior of the degree distribution is given by: 
\eq{
\lim_{t \rightarrow \infty} p_k(t) =2 \left(  \frac{1}{k+2} - \frac{2}{k+1} + \frac{1}{k} \right),
}
which simplifies to
\eq{
\lim_{t \rightarrow \infty} p_k(t) =   \frac{4}{k (k+1) (k+2)}.
}

As we mentioned previously, this asymptotic result was derived in \cite{rednerM1, rednerM2, rednerM4}. Also, for large values of $k$, this pertains to the $k^{-3}$ power law derived in \cite{barabSF2}.

\begin{figure}[t]
  \centering
  \includegraphics[width=1\columnwidth]{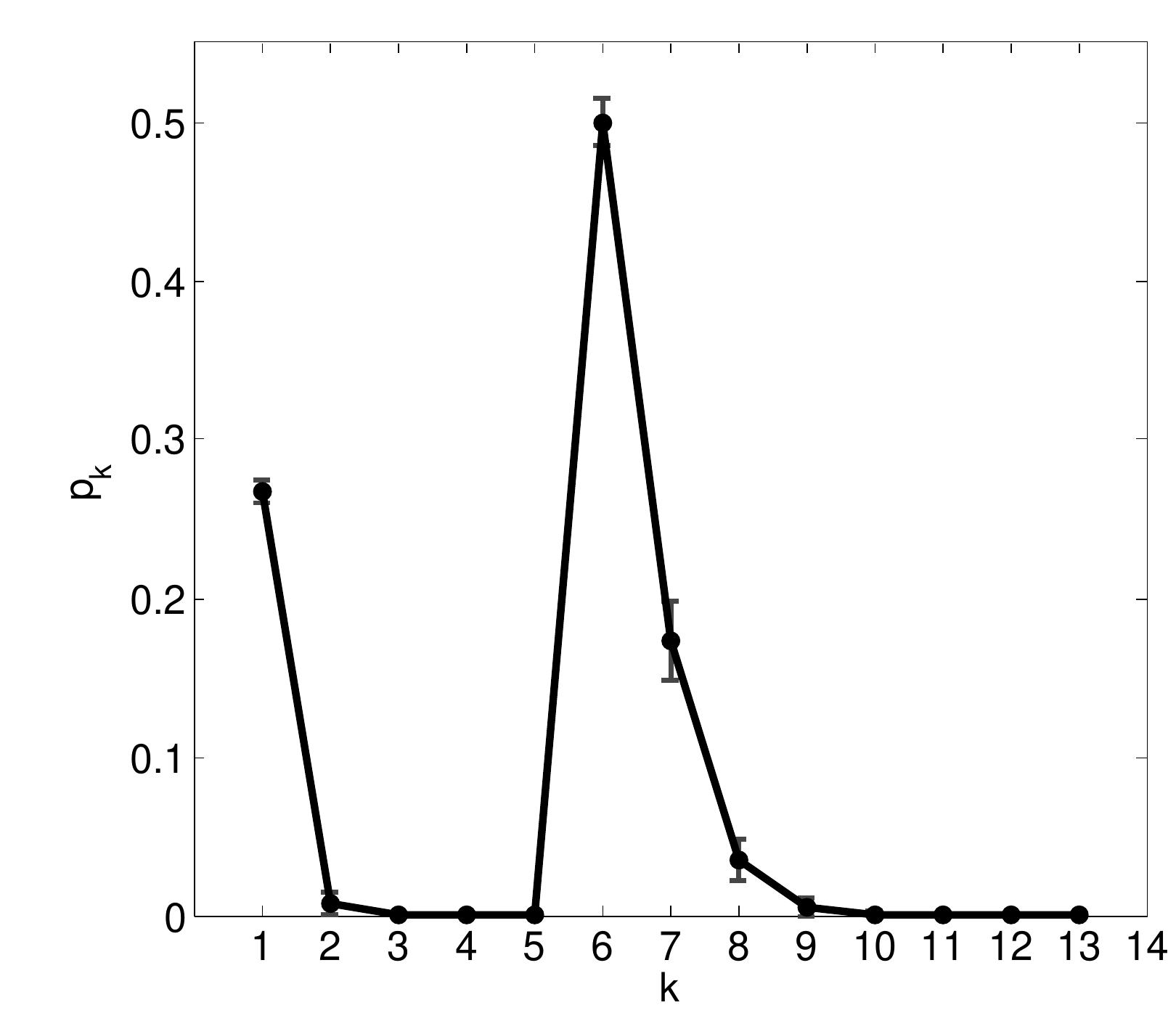}
  \caption[Figure 1]%
  {Degree distribution at time $t=20$ for a  6-regular ring of 50 nodes, subject to growth with preferential probabilities and single attachments.  The value of ${\alpha}$ is 1.  The error bars for 50 Monte~Carlo trials are plotted, along with the theoretical curve.  }
\label{fig_pref_single}
\end{figure}

Figure~\ref{fig_pref_single} shows $p_k(t)$ for $t=20$, for a  6-regular ring of total 50 nodes. It is seen in the figure that the second majority comprises of newcomers. The initial nodes of degree 6   who have received no new link  are  most frequent. Those who have received one new link and hence have degree 7  are  3rd-most  frequent.

\begin{figure}[t]
  \centering
  \includegraphics[width=1\columnwidth]{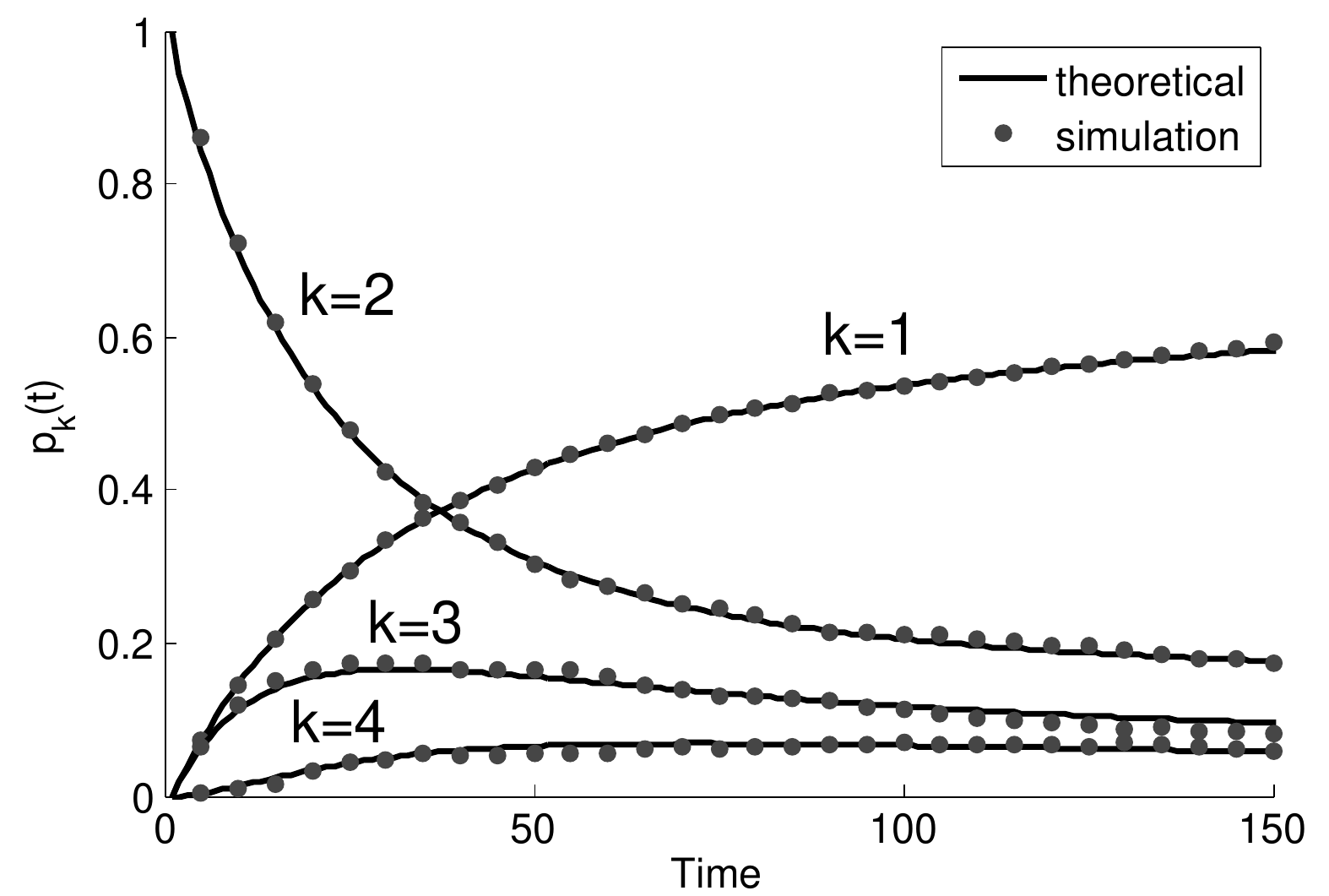}
  \caption[Figure 1]%
  {Growth under preferential probabilities and single attachments, on a ring (2-regular) of 30 nodes. The fractions of nodes having degree 2,3,4  are depicted in time. ${\alpha=1}$ and results are average over 30 Monte~Carlo trials.  }
\label{fig_pref_single_deg}
\end{figure}

Figure~\ref{fig_pref_single_deg}  illustrates the simulation results and theoretical predictions for  a ring~ (2-regular)  of 30 nodes, and $p_k(t)$ is presented as a function of time, for $k=2,3,4$. As seen in the figure, nodes  with degree 2 are mostly the initial nodes who have received no link from the newcomers, and their population diminishes as they receive new links and consequently turn into nodes of degree 3.

The mean and variance of simulations are presented  in Table~\ref{table_pref_single}.

\begin{table}[t]
\centering 
\begin{tabular}{| c |c |c |c|c|} 
\hline 
 \backslashbox{Time}{Degree} & $k=3$ & $k=4$ & $k=5$  & $k=6$\\ [0.5ex] 
\hline
t=5  & \dcell{0.0001\\ 0.0000} & \dcell{0.6839\\ 0.0006}& \dcell{0.1357\\ 0.0018}& \dcell{0.0131\\ 0.0002}\\     \hline
t=10  & \dcell{0.0006\\ 0.0000} & \dcell{0.4526\\ 0.0014}  & \dcell{0.1869\\ 0.0023}& \dcell{0.0420\\ 0.0007} \\ \hline
t=15  & \dcell{0.0018\\ 0.0000} & \dcell{0.3166\\ 0.0016} & \dcell{0.1819\\ 0.0022}& \dcell{0.0640\\ 0.0008}   \\ \hline
t=20  & \dcell{0.0036\\ 0.0001} & \dcell{0.2298\\ 0.0015} & \dcell{0.1722\\ 0.0017}& \dcell{0.0759\\ 0.0009}    \\ \hline
t=25  & \dcell{0.0056\\ 0.0001} & \dcell{0.1727\\ 0.0013} & \dcell{0.1527\\ 0.0014}& \dcell{0.0809\\ 0.0009}     \\ \hline
t=30  & \dcell{0.0078\\ 0.0001} & \dcell{0.1338 \\ 0.0011} & \dcell{0.1331\\ 0.0011}& \dcell{0.0819\\ 0.0008}   \\ \hline
t=35  & \dcell{0.0103\\ 0.0002} & \dcell{0.1059\\ 0.0010}& \dcell{0.1165\\ 0.0008}& \dcell{0.0782\\ 0.0009}    \\ \hline
t=40  & \dcell{0.0126\\ 0.0002} & \dcell{0.0860\\ 0.0008} & \dcell{0.1014\\ 0.0008}& \dcell{0.0746\\ 0.0009}   \\
 [1ex] 
\hline 
\end{tabular}
\caption{Preferential single attachment: the mean and variance (top and bottom row of each cell, respectively) of $p_k(t)$ for different instants of time and $k=3,4,5,6$. The substrate is a 4-regular ring of 20 nodes. the ensemble consists of 4000 Monte~Carlo trials.} 
\label{table_pref_single} 
\end{table}


\subsection{Multiple Connection}  \label{subsec:pref_multiple}

Now let us consider the preferential attachment scheme again, but this time, each new node attaches to $\beta$ existing nodes. At time $t$, the number of nodes will be ${N(0)+\alpha t}$. Also, at time $t$,  the sum of the degrees of all nodes (which equals twice the number of links) will be $ {N(0)\bar{k}_0+2\alpha \beta t}$. Each newly born node adds one to $N_{\beta}$ at that instant. Let us once again denote $N(0)\bar{k}_0$ by $\lambda$. Similar to~\eqref{Nkdot2}, the evolution of $N_k(t)$ is
\eq{
\dot{N_k}=\frac{\alpha \beta}{\sum_{\ell} \ell N_{\ell}} ((k-1)N_{k-1}-k N_k)+\alpha \delta_{k,\beta}.
}
Taking the Z-transform leads us to: 
\eqq{
\frac{\partial \psi}{\partial t}- \frac{\alpha \beta(z-1)}{\lambda + 2\alpha \beta t} 
\frac{\partial \psi}{\partial z}= \alpha z^{-\beta}
.}{PDE_pref_mult}
We solve this differential  equation via the method of characteristics in appendix~\ref{app:PDE_pref_multiple}. The result is:
\begin{align}
&\psi(z,t)=\psi_0\bigg((z-1) \sqrt{ \frac{\lambda+2\alpha \beta t}{\lambda}}+1\bigg) \nonumber \\
& -\frac{(z-1)^2 (\lambda + 2\alpha \beta t)}{\beta} 
\nonumber \\
&\times \bigg[ F(z)-F\bigg((z-1) \sqrt{ \frac{\lambda+2\alpha \beta t}{\lambda}}+1\bigg) \bigg]  ,
\label{psi_pref_mult2}
\end{align}
where the function $F(z)$ is defined as follows: 
\eq{
F(z) \stackrel{\text{def}}{=} \int^z \frac{x^{-\beta}}{(x-1)^3} dx
.}
Let us generalize~\eqref{c_def} and define the following:
\eq{
c \stackrel{\text{def}}{=} 1-\sqrt{\frac{\lambda}{\lambda+2\alpha \beta t}}
. } 
As above, this quantity is less than one and tends to one as ${t  \rightarrow \infty}$. Then~\eqref{psi_pref_mult2} is simplified to: 
\begin{align}
&\psi(z,t)=\psi_0\bigg(\frac{z-c}{1-c}\bigg) \nonumber \\
& -\frac{(z-1)^2 (\lambda + 2\alpha \beta t)}{\beta} \bigg[ F(z)-F\bigg( \frac{z-c}{1-c} \bigg) \bigg]  .
\label{psi_pref_mult3}
\end{align}
We invert the generating function in appendix~\ref{app: Z_pref_multiple}. Consequently,  for~$ t \geq 0$ and~$1 \leq  k \leq  N(0)$ we arrive at: 
\begin{align}
&N_k(t)= \sum_{\ell=1}^{k} N_{\ell}(0) (1-c)^{\ell} c^{k-\ell} \binom{k-1}{\ell-1} \nonumber \\
& +\frac{(N(0) \bar{k}_0 + 2\alpha \beta t)}{\beta} \frac{\beta(\beta+1)}{k(k+1)(k+2)} u(k-\beta) 
\nonumber \\
&- \frac{N(0) \bar{k}_0}{\beta}\sum_{\ell=\beta}^{k} \frac{\beta(\beta+1)}{\ell(\ell+1)(\ell+2)} (1-c)^{\ell} c^{k-\ell} \binom{k-1}{\ell-1},  
\label{Nk_pref_final_res}
\end{align}

Dividing this by the number of nodes at time $t$, which is equal to $N(0)+\alpha t$, yields the degree distribution at time $t$. As above, let us denote the degree distribution of the initial graph by $n_k$.  The final result for the degree distribution   for~$ t \geq 0$ and~$1 \leq  k \leq  N(0)$ is as follows: 
\begin{align}
&p_k(t)= \frac{N(0)}{N(0)+\alpha t}\sum_{\ell}  n_{\ell}(1-c)^{\ell} c^{k-\ell} \binom{k-1}{\ell-1}
\nonumber \\
& +\frac{(N(0) \bar{k}_0 + 2\alpha \beta t)}{N(0)+\alpha t} \frac{(\beta+1)}{k(k+1)(k+2)} u(k-\beta) 
\nonumber \\
&-\frac{N(0) \bar{k}_0}{N(0)+\alpha t}\sum_{\ell=\beta}^{k} \frac{(\beta+1)}{\ell(\ell+1)(\ell+2)} (1-c)^{\ell} c^{k-\ell} \binom{k-1}{\ell-1}.
\label{pk_pref_multiple}
\end{align}
The equivalence of this result for the special case of~$\beta=1$ with~\eqref{pk_pref_single} is proved in appendix~\ref{app:equivalence}. 

Now let us focus on the long time behavior, when $t \rightarrow \infty$, we have: 
\eq{
\begin{cases}
\displaystyle \lim_{t \rightarrow \infty}  \frac{N(0)}{N(0)+\alpha t}=0 \\ \\
\displaystyle \lim_{t \rightarrow \infty} \frac{N(0) \bar{k}_0+2\alpha \beta t}{N(0)+\alpha t} =2\beta \\ \\
\displaystyle \lim_{t \rightarrow \infty} c = 1.
\end{cases}
}
Using these values, the asymptotic degree distribution is obtained: 
\eq{
\lim_{t \rightarrow \infty} p_k(t)= \frac{2\beta(\beta+1)}{k(k+1)(k+2)} u(k-\beta),
}
which matches~\eqref{bol_res_1}.

\begin{figure}[t]
  \centering
  \includegraphics[width=1\columnwidth]{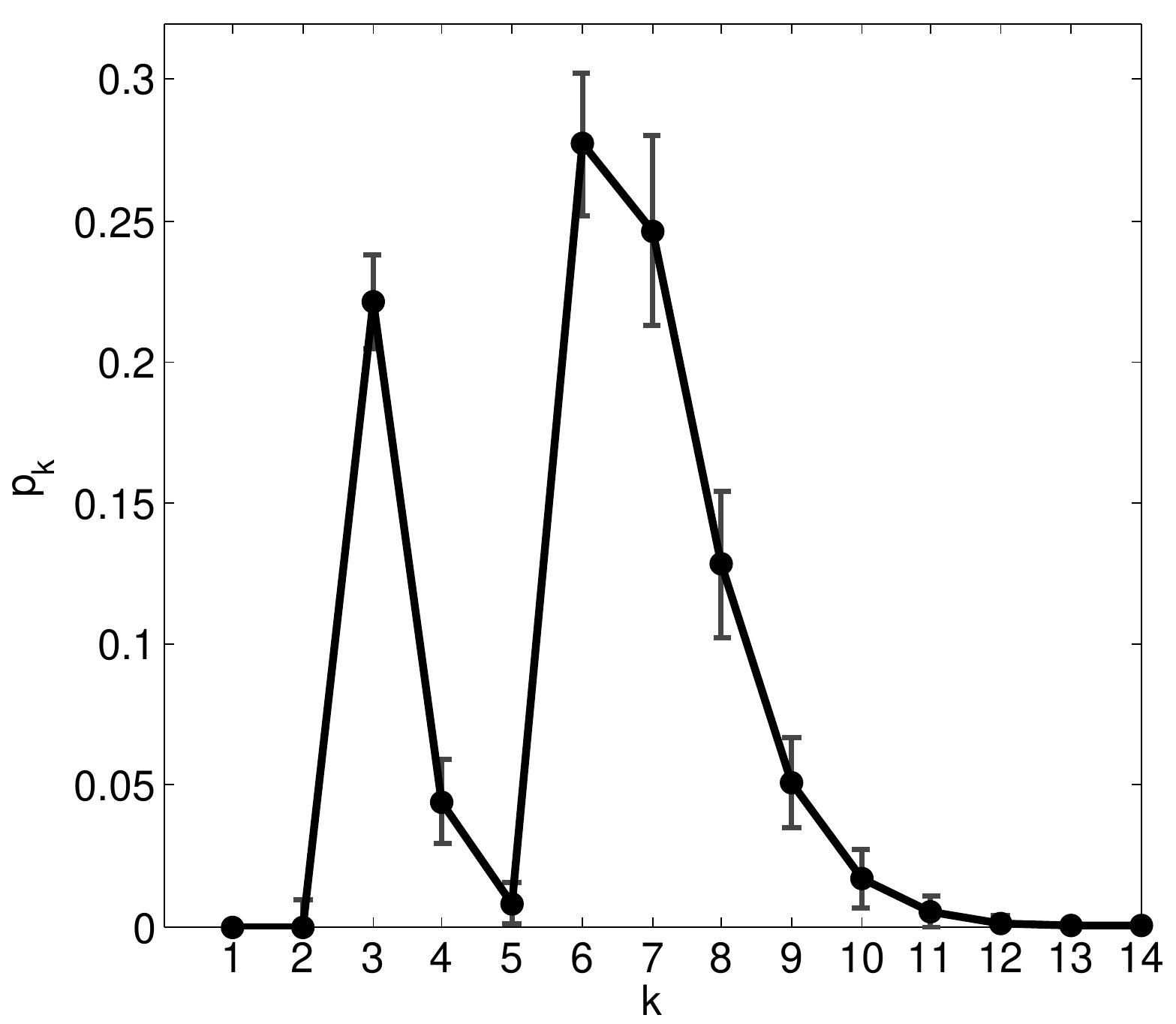}
  \caption[Figure 1]%
  {Degree distribution at time $t=20$ for a  6-regular ring of 50 nodes, subject to growth with preferential probabilities and multiple attachments, with ${\beta=3}$. The value of $\alpha$ is 1. The error bars for 50 Monte~Carlo trials are plotted, along with the theoretical curve.}
\label{fig_pref_multiple}
\end{figure}

Figure~\ref{fig_pref_multiple} shows $p_k(t)$ for $t=20$, for a  6-regular ring of total 50 nodes. The value of $\beta$ is 3.  As seen in the figure, the left-most  peak belongs to those with degree $\beta$, that are the newly added nodes. Those with degree 6 and 7, that are mostly  the initial nodes who have received zero and one links from the newcomers respectively,  are first and second most frequent, as seen in the graph.

\begin{figure}[t]
  \centering
  \includegraphics[width=1\columnwidth]{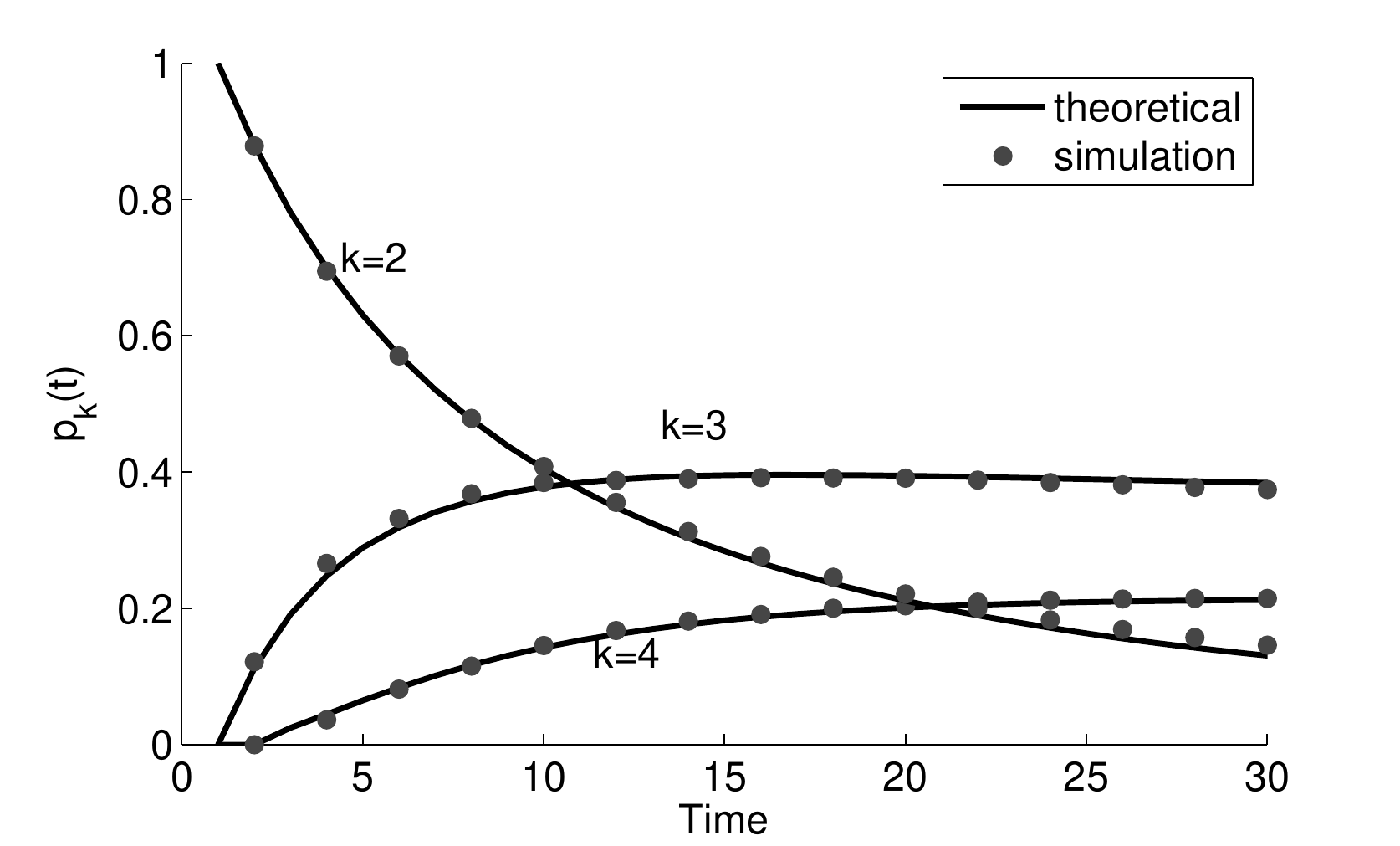}
  \caption[Figure 1]%
  {Growth under preferential probabilities and multiple attachments, on a ring (2-regular) of 30 nodes. The fractions of nodes having degree 2,3,4  are depicted in time. The value of $\beta$ is 3 and the value of $\alpha$ is 1. The results are average over 30 Monte~Carlo trials. }
\label{fig_pref_multiple_deg}
\end{figure}

Figure~\ref{fig_pref_multiple_deg}  is a depiction the simulation results and theoretical predictions for  a ring~ (2-regular)  of 30 nodes, and $p_k(t)$ is presented as a function of time, for $k=2,3,4$.  The value of $\beta$ is 3. It is observable  in the figure that the network keeps losing nodes of degree 2 (the initial nodes) as they receive links from the newcomers.

The mean and variance of simulations are presented  in Table~\ref{table_pref_mult}.

\begin{figure}[t]
  \centering
  \includegraphics[width=1\columnwidth]{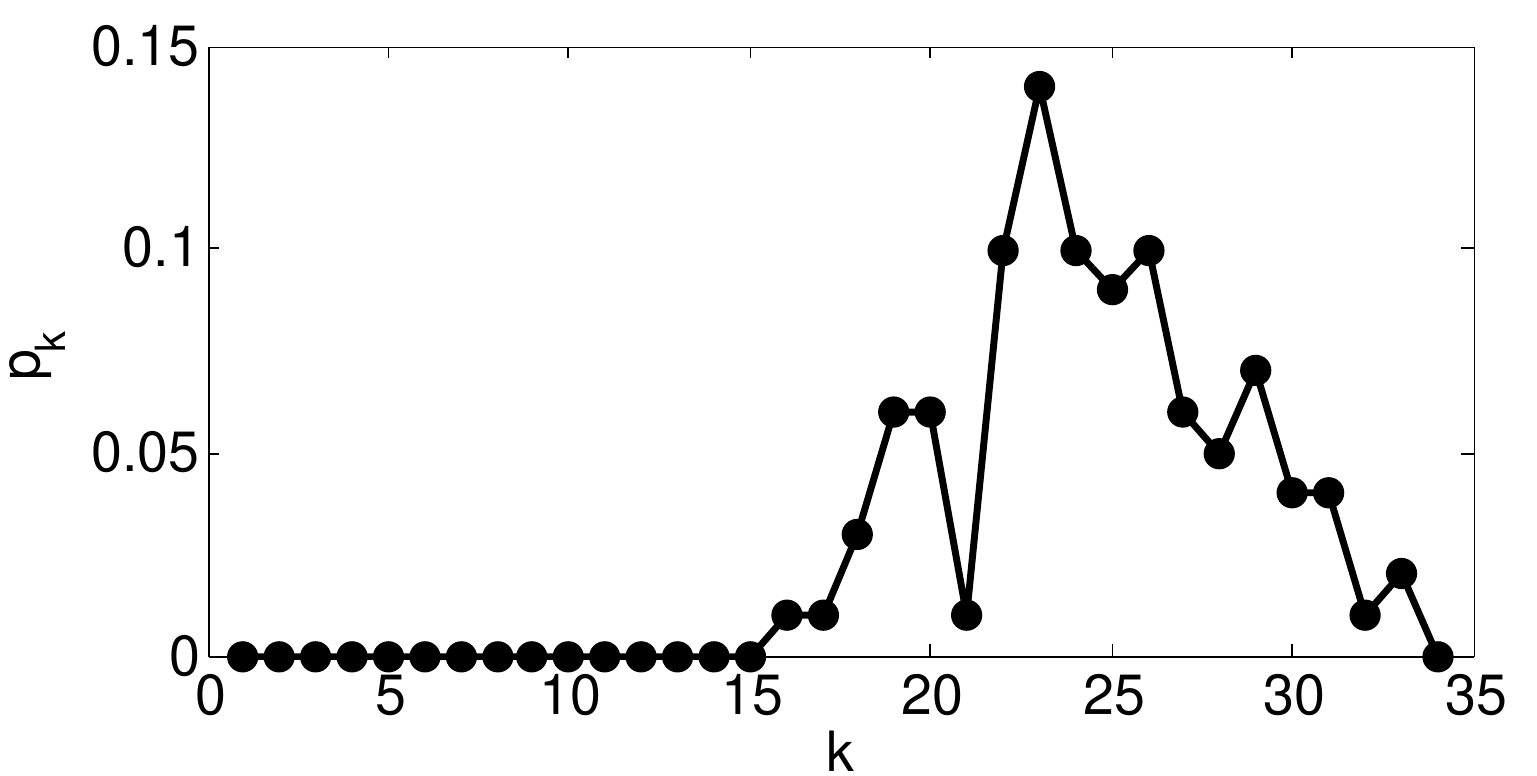}
  \caption[Figure 1]%
  {The degree distribution for a Small-world graph which is used as the substrate for preferential growth. It is constructed by adding random long-range links to a 6-regular ring. Each long-range link is formed with probability 0.2.}
\label{SW_0}
\end{figure}

\begin{figure}[t]
  \centering
  \includegraphics[width=1\columnwidth]{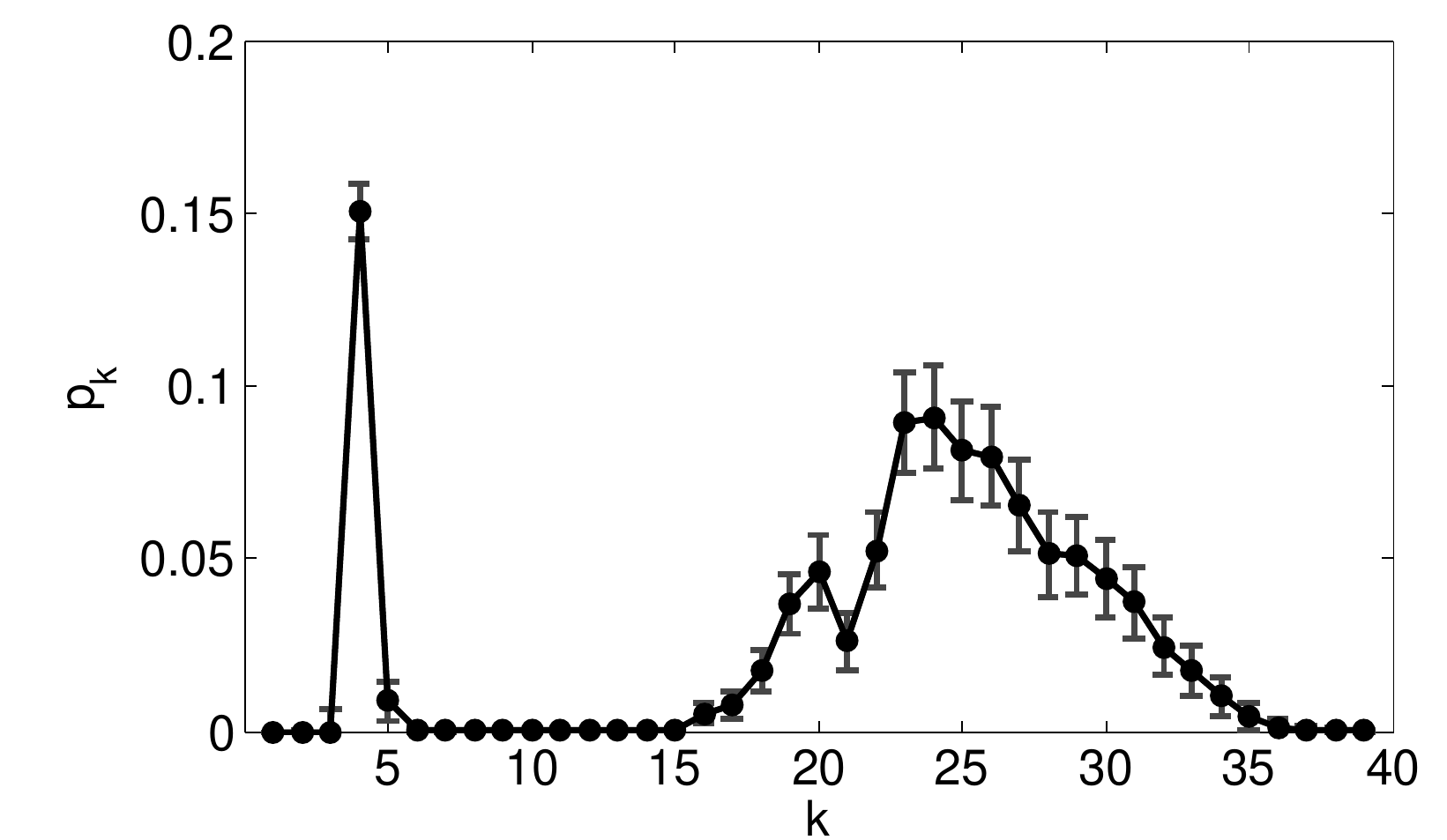}
  \caption[Figure 1]%
  {Degree distribution at time $t=20$ for preferential growth with $\beta=4$ and $\alpha=1$  over the Small-world graph whose degree distribution is depicted in Figure~\ref{SW_0}. Theoretical prediction is illustrated along with simulation error bars. The error bars are for 50 Monte-Carlo trials.}
\label{SW_20}
\end{figure}

\begin{figure}[t]
  \centering
  \includegraphics[width=1\columnwidth]{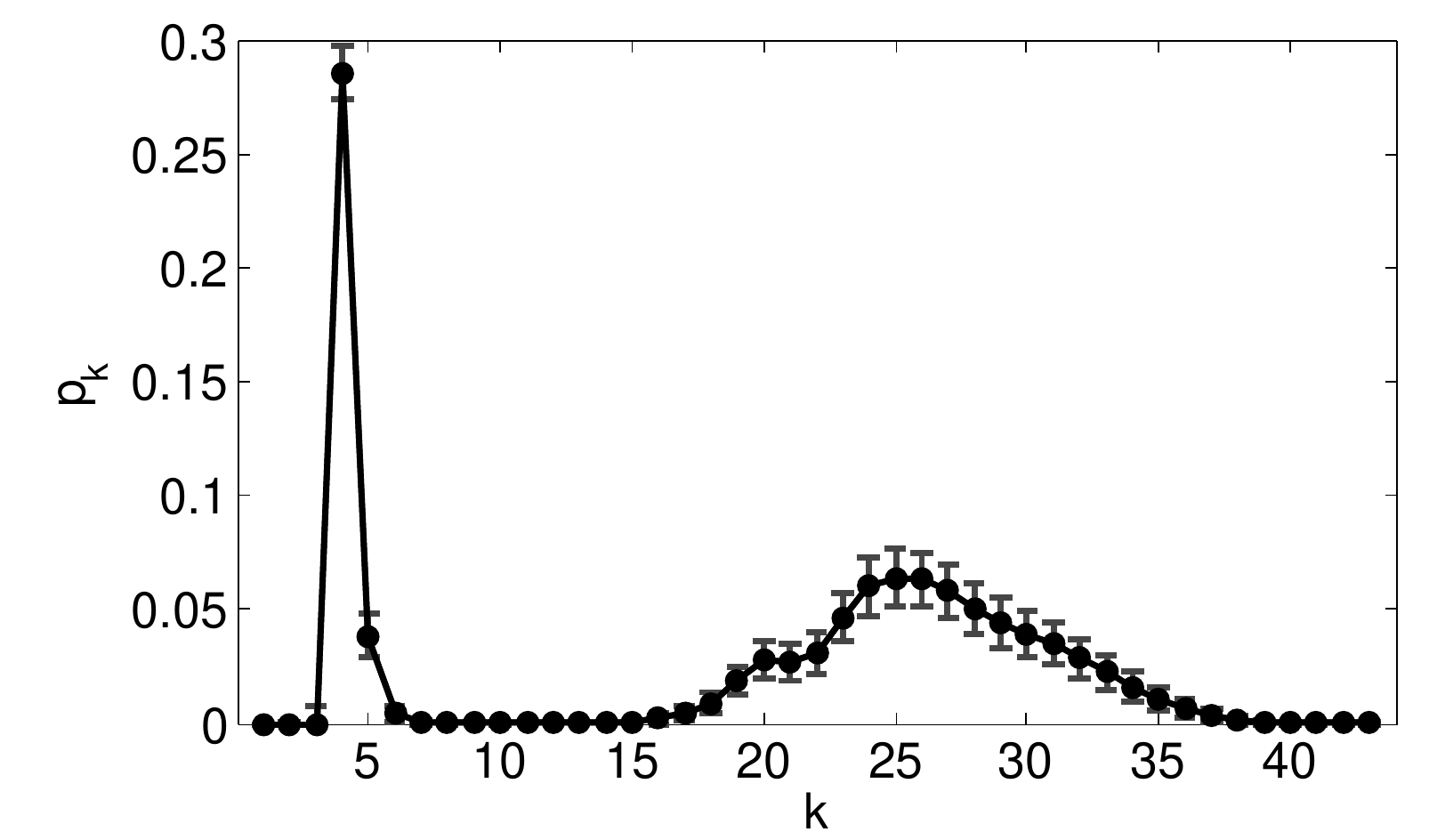}
  \caption[Figure 1]%
  {Degree distribution at time $t=50$ for preferential growth with $\beta=4$ and $\alpha=1$  over the Small-world graph whose degree distribution is depicted in Figure~\ref{SW_0}. Theoretical prediction is illustrated along with simulation error bars. The error bars are for 50 Monte-Carlo trials.}
\label{SW_50}
\end{figure}

\begin{figure}[t]
  \centering
  \includegraphics[width=1\columnwidth]{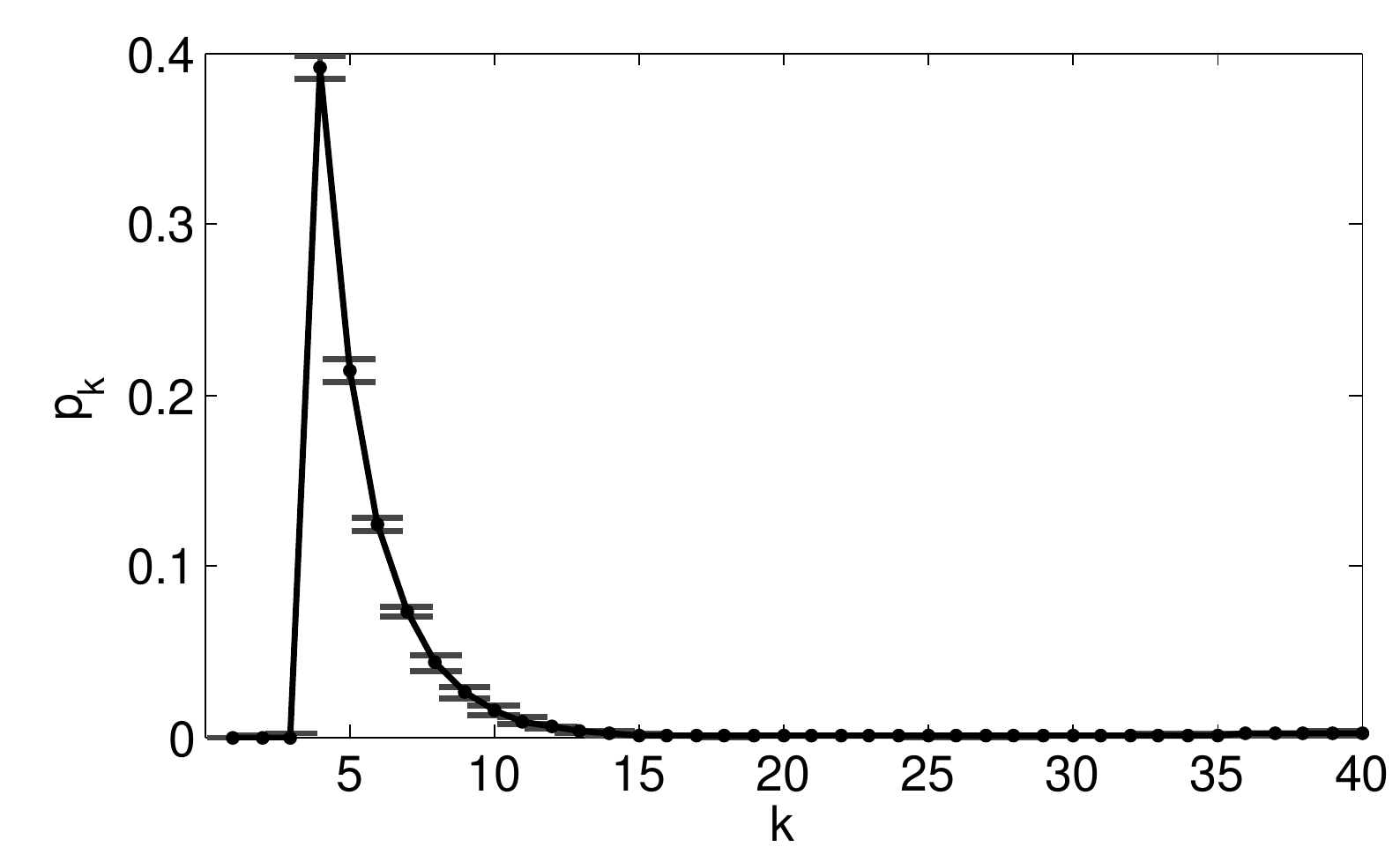}
  \caption[Figure 1]%
  {Degree distribution at time $t=1000$ for preferential growth with $\beta=4$ and $\alpha=1$  over the Small-world graph whose degree distribution is depicted in Figure~\ref{SW_0}. Theoretical prediction is illustrated along with simulation error bars. The error bars are for 50 Monte-Carlo trials. It can be seen that at this time, the effects of the initial graph have almost entirely vanished and the degree distribution resembles a pure power-law curve. Also note that the fraction of nodes with degree less than $\beta$ is close almost zero.}
\label{SW_500}
\end{figure}

Let us also compare theoretical predictions and simulation results for a graph which is not regular, which means that the degree of all nodes are not necessarily the same~\footnote{There is no technical reason why we used regular graphs for simulations. The initial substrate of network growth can have arbitrary structure. We  presented simulation results for a Small-world graph, to illustrate the point that regularity of the initial graph has no role in the validity of the results}. We use a Small-world graph~\cite{watts1}. To construct the graph, we take a 6-regular ring of 100 nodes and add random links between nodes that are not connected. Each link is formed with probability 0.2. The degree distribution of the resulting substrate is given in Figure~\ref{SW_0}. Then the preferential growth on the network begins with $\beta=4$. The degree distribution at $t=20$ is illustrated in Figure~\ref{SW_20}, and the degree distribution at time $t=50$ is presented in Figure~\ref{SW_50}.   Figure~\ref{SW_500} is a depiction of the degree distribution at $t=1000$, which is close to the `long time limit'. The effect of the initial graph has almost disappeared  and the distribution is close to the power-law form. Also note that  the fraction of nodes with degree smaller  than $\beta$  tends to  zero in this time regime.

\begin{table}[t]
\centering 
\begin{tabular}{| c |c |c |c|c|} 
\hline 
 \backslashbox{Time}{Degree} & $k=3$ & $k=4$ & $k=5$  & $k=6$\\ [0.5ex] 
\hline
t=5  & \dcell{0.1390\\ 0.0004} & \dcell{0.5165\\ 0.0013}& \dcell{0.2462\\ 0.0022}& \dcell{0.0719\\ 0.0007}\\     \hline
t=10  & \dcell{0.2232\\ 0.0006} & \dcell{0.3092\\ 0.0016}  & \dcell{0.2410\\ 0.0020}& \dcell{0.1317\\ 0.0012} \\ \hline
t=15  & \dcell{0.2681\\ 0.0007} & \dcell{0.2337\\ 0.0014} & \dcell{0.2016\\ 0.0016}& \dcell{0.1386\\ 0.0012}   \\ \hline
t=20  & \dcell{0.2954\\ 0.0007} & \dcell{0.2025\\ 0.0011} & \dcell{0.1708\\ 0.0011}& \dcell{0.1301\\ 0.0010}    \\ \hline
t=25  & \dcell{0.3135\\ 0.0007} & \dcell{0.1887\\ 0.0009} & \dcell{0.1501\\ 0.0009}& \dcell{0.1188\\ 0.0007}     \\ \hline
t=30  & \dcell{0.3263\\ 0.0006} & \dcell{0.1826 \\ 0.0007} & \dcell{0.1364\\ 0.0008}& \dcell{0.1086\\ 0.0006}   \\ \hline
t=35  & \dcell{0.3359\\ 0.0006} & \dcell{0.1800\\ 0.0006}& \dcell{0.1273\\ 0.0006}& \dcell{0.1003\\ 0.0006}    \\ \hline
t=40  & \dcell{0.3432\\ 0.0006} & \dcell{0.1792\\ 0.0005} & \dcell{0.1212\\ 0.0005}& \dcell{0.0936\\ 0.0004}   \\
 [1ex] 
\hline 
\end{tabular}
\caption{Preferential multiple attachment: the mean and variance (top and bottom row of each cell, respectively) of $p_k(t)$ for different instants of time and $k=3,4,5,6$. The substrate is a 4-regular ring of 20 nodes.The value of~$\beta$ is 3. The ensemble consists of 4000 Monte~Carlo trials.} 
\label{table_pref_mult} 
\end{table}

\section{Summary and Future Work} \label{sec:conclusion}

Previous work in the literature of network growth models mainly focus on the degree distribution of the graph in the asymptotic limit, that is, when the number of nodes tends to infinity and the effect of initial conditions can be neglected. In this contribution we found time-dependent expressions for the expected degree distribution, which depend explicitly on the degree distribution of the initial graph. We considered two growth schemes. One in which new nodes choose from existing nodes uniformly at random, and then connect to them, and the other where these probabilities are proportional to degrees. Uniform and multiple attachments for the newly-born nodes are considered separately for both cases. Simulation results were accompanying theoretical predictions for each case. 

One possible extension of the results presented in this work would be as follows. Suppose a given graph is subject to growth. The current state of the graph is known, and the growth mechanism can be approximated to be uniformly at random or be preferential attachment. Suppose quite on the contrary to the previous work in the literature, we are interested in the short-time behavior of the degree distribution. Then one could employ the results in this work, and expand the expressions in the vicinity of the initial condition up to arbitrary order of ${(\alpha t)/N(0)}$, and find the degree distribution perturbatively, to arbitrary precision. 

Our analysis focuses on the expected degree distribution. Due to the random nature of the growth process, $p_k(t)$ has a distribution of its own, whose mean value is presented in this work. One can also focus on the variance, or other statistical properties,  of this distribution.

\section{Acknowledgment}
This work was funded in part by the Natural Sciences and Engineering Research Council of Canada.


\appendix

\section{Solving Equation (\ref{psi_PDE_1}) for Uniform Single Attachment} \label{app: solve_PDE_unif_single} 
The equation is repeated here for easy reference: 
\eq{
\frac{\partial \psi(z,t)}{\partial t}-\frac{\alpha (z^{-1}-1)}{N(0)+\alpha t}\psi(z,t)= \frac{\alpha}{z},
}
with the following general form of a first order linear equation in time domain: 
\eq{
\dot{\psi}+p(z,t)\psi =q(z,t),
}
Multiply both sides by an unknown integrating factor ${\mu(z,t)}$
to make both sides equal to ${\frac{\partial}{\partial t}\big[ \mu(z,t) \psi(z,t)\big]}$. Then ${\mu(z,t)}$ is found to be $\exp[\int p(z,t)dt]$. Thus the final solution becomes: 
\eqq{
\psi(z,t)=\frac{1}{\mu(z,t)} \bigg[ \int \mu(z,t) q(z,t) dt + C(z) \bigg]
,}{sol1}
where $C(z)$ depends on the initial conditions. In our problem,  the initial degree distribution is given and it will be used to determine~$C(z)$. For~${\mu(z,t)}$ we have:
\begin{align}
\mu(t,z)&= \exp \bigg[-\int \frac{\alpha (z^{-1}-1)}{N(0)+\alpha t} dt\bigg] \nonumber \\
&= \exp \bigg[ (1-z^{-1}) \ln [N(0)+\alpha t] \bigg] \nonumber \\
&= [N(0)+\alpha t]^{(1-z^{-1})}
.
\end{align}
Using this and~\eqref{sol1} , we find $\psi(z,t)$ as follows: 
\eq{
\displaystyle \psi(z,t)=\frac{\bigg[ \int \frac{\alpha}{z}[N(0)+\alpha t]^{(1-z^{-1})}dt + C(z)   \bigg]
}{[N(0)+\alpha t]^{(1-z^{-1})} }
.}
Note that the constant depends on $z$. Carrying out the integral, we arrive at: 
\eqq{
\psi(z,t)=\frac{N(0)+\alpha t}{z(2-\frac{1}{z})}+C(z)(N(0)+\alpha t)^{\frac{1}{z}-1}
.}{psi_unif_1}
The constant $C(z)$ is obtained by looking at $t=0$, when the following holds: 
\eq{
\psi(z,0)=
\frac{N(0)}{z(2-\frac{1}{z})}+C(z) N(0)^{\frac{1}{z}-1}
.}
Solving this for $C(z)$ and plugging the result in~\eqref{psi_unif_1}  
yields
\begin{align}
\psi(z,t)&=\frac{N(0)+\alpha t}{z(2-\frac{1}{z})}+\frac{N(0)}{N(0)+\alpha t} \left[\psi(z,0) (1+\frac{\alpha t}{N(0)})^{\frac{1}{z}}\right] \nonumber \\
&-
\frac{N(0)^2}{N(0)+\alpha t} \frac{1}{z(2-\frac{1}{z})}\left(1+\frac{\alpha t}{N(0)}\right)^{\frac{1}{z}}.
\label{psi_unif_Z}
\end{align}


\section{Solving equation~\eqref{psi_PDE_2} for Uniform Multiple Attachment} \label{app: solve_PDE_unif_multiple} 
The following differential equation must be solved:  
\eq{
\frac{\partial \psi(z,t)}{\partial t}-\frac{\beta \alpha (z^{-1}-1)}{N(0)+\alpha t}\psi(z,t)= \frac{\alpha}{z^\beta}.
}
The integration factor is: 
\begin{align}
\mu(t,z)&= \exp \bigg[-\int \frac{\beta \alpha (z^{-1}-1)}{N(0)+\alpha t} dt\bigg] \nonumber \\
&= \exp \bigg[ \beta(1-z^{-1}) \ln [N(0)+\alpha t] \bigg] \nonumber \\
&= [N(0)+\alpha t]^{\beta(1-z^{-1})}
.
\end{align}
And from~\eqref{sol1} , $\psi(z,t)$ is obtained: 
\eq{
\displaystyle \psi(z,t)=\frac{\bigg[ \int \frac{ \alpha}{z^{\beta}}[N(0)+\alpha t]^{\beta(1-z^{-1})}dt + C(z)   \bigg]
}{[N(0)+\alpha t]^{\beta(1-z^{-1})} }
.}
After integration, we get: 
\eqq{
\psi(z,t)=\frac{1}{z^{\beta}}\frac{N(0)+\alpha t}{1+\beta(1-z^{-1})}+
\frac{C(z)}{\big[N(0)+\alpha t\big]^{\beta(1-\frac{1}{z})}}
.}{psi_unif_2}
Setting $t=0$, we find $C(z)$:
\eq{
C(z)=\bigg[ \psi(z,0)-\frac{N(0)z^{-\beta}}{1+\beta(1-z^{-1})} \bigg] N(0)^{\beta(1-z^{-1})}
.}
Substituting this $C(z)$ in~\eqref{psi_unif_2} yields:
\begin{align}
\psi(z,t)&=\frac{1}{z^\beta} \frac{N(0)+\alpha t}{1+\beta(1-z^{-1})} \nonumber \\
&+ \psi(z,0) \bigg[\frac{N(0)}{N(0)+\alpha t} \bigg]^{\beta(1-z^{-1})} \nonumber \\
&- \frac{N(0)}{z^\beta \big[ 1+\beta(1-z^{-1})\big] }  \bigg[\frac{N(0)}{N(0)+\alpha t} \bigg]^{\beta(1-z^{-1})}
.
\end{align}


\section{Taking the Inverse Z-transform of~\eqref{psi_unif_Z_2}} \label{app: Z_unif_multiple}

The following function of $z$ must be inverted:
\begin{align}
\psi(z,t)&=\frac{1}{z^\beta} \frac{N(0)+\alpha t}{1+\beta(1-z^{-1})} \nonumber \\
&+ \psi(z,0) \bigg[\frac{N(0)}{N(0)+\alpha t} \bigg]^{\beta(1-z^{-1})} \nonumber \\
&- \frac{N(0)}{z^\beta \big[ 1+\beta(1-z^{-1})\big] }  \bigg[\frac{N(0)}{N(0)+\alpha t} \bigg]^{\beta(1-z^{-1})}
.
\end{align}
Now we must invert this, term by term. This is done in appendix~\ref{app: Z_unif_multiple}
First, note that: 
\eq{
\frac{1}{z^\beta \big[ 1+\beta (1-z^{-1}) \big] } = (1+\beta)^{-1} \frac{1}{z^{\beta-1}
\big[ z-\frac{\beta}{\beta+1}\big]}
.}
Now, by definition, the inverse Z-transform is given by: 
\begin{align}
\displaystyle N_k(t)&= 
\frac{1}{2\pi i} \oint \psi(z,t) z^{k-1} dz  \nonumber \\
&=(1+\beta)^{-1} \oint \frac{z^{k-\beta}}{z-\frac{\beta}{\beta+1}} dz
.
\end{align}
In order to perform the integration, one must find the residuals of $\frac{z^{k-\beta}}{z-\frac{\beta}{\beta+1}}$~\cite{morse,resid1}. For this purpose, we have to assume three distinct cases: 
\begin{itemize}
\item If $k=\beta$, then we are looking for the residuals of $\frac{1}{z-\frac{\beta}{\beta+1}}$ which is unity. 
\item If $k>\beta$, then the only pole is at $z=\frac{\beta}{\beta+1}$ and the residual becomes 
$\big[ \frac{\beta}{\beta+1} \big]^{k-\beta}$. 
\item If $k<\beta$, then $z=0$ is also a pole, and is of order $\beta-k$. The residual at this pole is equal to the following: 
\eq{
\frac{1}{(\beta-k-1)!}\frac{d^{(\beta-k-1)}}{dt^{(\beta-k-1)}}  \frac{1}{z-\frac{\beta}{\beta+1}} \bigg|_{z=0} = - \bigg[ \frac{\beta+1}{\beta} \bigg]^{\beta-k}
.}
The residual for the pole at $z=\frac{\beta}{\beta+1}$ is positive the same value, so they cancel out. 
\end{itemize}
Putting these three together, we find the inverse transform: 
\eq{
\frac{1}{z^{\beta-1}
\big[ z-\frac{\beta}{\beta+1}\big]}  \xrightarrow{\mathcal{Z}^{-1}}  
\bigg[ \frac{\beta}{\beta+1} \bigg]^{k-\beta} u(k-\beta)
.}
We have inverted the first term of~\eqref{psi_unif_Z_2}. Another inverse transform that we need is the following. 
\eq{
\bigg[\frac{N(0)}{N(0)+\alpha t}\bigg]^{-\beta z^{-1}} 
 \xrightarrow{\mathcal{Z}^{-1}}  
 \frac{\bigg[ \beta \ln \bigg( 1+\frac{\alpha t}{N(0)} \bigg) \bigg]^k}{k!}.
}
Using these two results, the inverse transform of~\eqref{psi_unif_Z_2} becomes: 
\begin{align}
&N_k(t)=
\frac{N(0)+\alpha t}{\beta} \left( \frac{\beta}{\beta+1} \right)^{k-\beta+1} u(k-\beta)
\nonumber \\
&+ \left[ \frac{N(0)^{\beta+1}}{(N(0)+\alpha t)^{\beta}} \right]    \left\{ n_k * \frac{\left[\beta  \ln \left( 1+\frac{\alpha t}{N(0)} \right) \right]^k}{k!} \right \}
\nonumber \\
&-\left[ \frac{N(0)^{\beta+1}}{\beta(N(0)+\alpha t)^{\beta}} \right] \times \nonumber \\
&\left\{ \left( \frac{\beta}{\beta+1} \right)^{k-\beta+1} u(k-\beta)  * \frac{\left[\beta  \ln \left( 1+\frac{\alpha t}{N(0)} \right) \right]^k}{k!} \right\},
\end{align}


\section{Method of Characteristics} \label{app:PDE}
Here we illustrate the basic procedure with an example. Consider the following partial differential equation for a function $\psi(x,y)$: 
\eq{
x^2 \frac{\partial \psi(x,y)}{\partial x} + y^3 \frac{\partial \psi(x,y)}{\partial y} = \psi^4(x,y)
~~~,x\geq1, y\geq0.}
First we solve the following system of equations: 
\eq{
\frac{dx}{x^2}=\frac{dy}{y^3}=\frac{d\psi}{\psi^4}
.}
The first equation is solved as follows: 
\eq{
\frac{dx}{x^2}=\frac{dy}{y^3} \Longrightarrow \frac{1}{x}-\frac{1}{2y^2}=C_1
,}
where $C_1$ is a constant. For the next equation we have (we arbitrarily pick one of the two remaining equations): 
\eq{
\frac{d\psi}{\psi^4}=\frac{dy}{y^3} \Longrightarrow \frac{1}{3 \psi^3}-\frac{1}{2y^2}=C_2
.}
Then the solution is of the form: 
\eq{ 
F(C_1)+G(C_2)=0
,}
for any differentiable function $F(\cdot)$ and $G(\cdot)$. This is equivalent to: 
\eq{
C_2=H (C_1)
,}
for an arbitrary function $H(\cdot)$. This means that: 
\eq{
\frac{1}{3 \psi^3}-\frac{1}{2y^2}= H\big( \frac{1}{x}-\frac{1}{2y^2} \big)
.}
This can be simplified to give: 
\eq{
\frac{1}{ \psi^3}=\frac{3}{2y^2}+3 H \big( \frac{1}{x}-\frac{1}{2y^2} \big)
,}
Denoting $3H(\cdot)$ by $\Phi(\cdot)$, the final solution can be expressed in the following general form for any differentiable function $\Phi(\cdot)$: 
\eqq{
\psi(x,y)=\bigg[ \frac{3}{2y^2}+\Phi \big( \frac{1}{x}-\frac{1}{2y^2} \big) \bigg]^{-\frac{1}{3}}
.}{example_1}
The function $\Phi(\cdot)$ is uniquely determined using the boundary conditions. Suppose we have the following information for the $x=1$ boundary: 
\eq{
\psi(1,y)= e^y
.}
Plugging $x=1$ in~\eqref{example_1} yields:
\eq{
\psi(1,y)=\bigg[ \frac{3}{2y^2}+\Phi \big( 1-\frac{1}{2y^2} \big) \bigg]^{-\frac{1}{3}} =e^y
.
}
Simplifying this, we get: 
\eq{
\Phi \big( 1-\frac{1}{2y^2} \big)= e^{-3y}-\frac{3}{2y^2}
.}
If we denote $1-\frac{1}{2y^2}$ by $X$, then $y=\sqrt{\frac{1}{2(1-X)}}$. So we arrive at: 
\eq{
\Phi (X)= \exp \bigg[-3\sqrt{\frac{1}{2(1-X)}}\bigg]-3(1-X)
.}
This $\Phi(\cdot)$ provides the unique solution of the form~\eqref{example_1},  given the boundary condition $\psi(1,y)= e^y$. This concludes the example.


\section{Solving Equation~\eqref{diff_eq_pref} Using the Method Of Characteristics} \label{app: PDE_pref_single}

We seek the solution of the following partial differential equation:
\eq{
\frac{\partial \psi}{\partial t}- \frac{\alpha (z-1)}{\lambda + 2\alpha t} 
\frac{\partial \psi}{\partial z}= \alpha z^{-1}
.}
As described in appendix~\ref{app:PDE}, we must solve the following system of equations:
\eq{
\displaystyle \frac{dt}{1}=\frac{dz}{\big[- \frac{\alpha (z-1)}{\lambda + 2\alpha t}  \big]}
=\frac{d\psi}{\alpha z^{-1}}
.
}
The simplified version of the first equation is the following: 
\eq{
\displaystyle \frac{dt}{\lambda + 2\alpha t}=\frac{-dz}{ \alpha (z-1)}
.}
Whose solution is the following:
\eqq{
(z-1)^2 (\lambda + 2\alpha t)= C
,}{C1}
where $C$ is a constant. The second equation is: 
\eq{
\frac{d\psi}{\alpha z^{-1}}= \frac{-dz}{\alpha (z-1)}(\lambda + 2\alpha t)
.}
Replacing ${(\lambda + 2\alpha t)}$ by $C/(z-1)^2$ yields the following: 
\eq{
d\psi = \frac{-C dz}{z(z-1)^3}
.}
Integrating both sides, gives: 
\eq{
\psi-C\bigg[ \frac{-1}{z-1}+\frac{1}{2(z-1)^2}+\ln\frac{z}{z-1} \bigg]=C'
,}
where $C'$ is another constant. Following the lines of the example, we know that the solution has the following form for some function $\Phi(\cdot)$ which must be determined from the initial conditions: 
\eq{
\psi-C\bigg[ \frac{-1}{z-1}+\frac{1}{2(z-1)^2}+\ln\frac{z}{z-1} \bigg]=
\Phi\bigg[ (z-1)^2 (\lambda + 2\alpha t) \bigg] 
.}
Replacing $C$ from~\eqref{C1} we get: 
\begin{align}
&\Phi\big[ (z-1)^2 (\lambda + 2\alpha t) \big] 
=\psi(z,t) \nonumber \\
&- (\lambda + 2 \alpha t) \bigg[ -(z-1)+\frac{1}{2}+(z-1)^2 \ln\frac{z}{z-1} \bigg]
. \label{psi_pref_1}
\end{align}
We must determine the function $\Phi(\cdot)$. Suppose the $N_k$s at the outset are known, so that for the initial graph we know $\psi(z,0)$. Let us denote it by $\psi_0(z)$. Setting $t=0$ in~\eqref{psi_pref_1} we get: 
\begin{align}
&\Phi\big[ (z-1)^2 \lambda \big] 
=\psi_0(z) \nonumber \\
&- \lambda \bigg[ -(z-1)+\frac{1}{2}+(z-1)^2 \ln\frac{z}{z-1} \bigg]
. 
\end{align}
This helps us determine the function $\Phi(X)$. Denoting ${(z-1)^2 \lambda}$ by $X$,  we get: 
\eq{
\Phi(X) 
=\psi_0(\sqrt{\frac{X}{\lambda}}+1) \nonumber \\
- \lambda \bigg[ -\sqrt{\frac{X}{\lambda}}+\frac{1}{2}+\frac{X}{\lambda} 
\ln\frac{\sqrt{X}+\sqrt{\lambda}}{\sqrt{\lambda}} \bigg]
.}
Since in~\eqref{psi_pref_1} we have ${[(z-1)^2 \lambda]}$ as $X$, let us explicitly find ${\Phi((z-1)^2 \lambda)}$. We have:
\begin{align}
&\Phi\big[ (z-1)^2 \lambda \big] 
=\psi_0\bigg[(z-1)\sqrt{\frac{\lambda+2\alpha t}{\lambda}}+1\bigg]  \nonumber \\
&- \lambda \bigg[ -(z-1)\sqrt{\frac{\lambda+2\alpha t}{\lambda}}+\frac{1}{2}  \nonumber \\
&+(z-1)^2 \frac{\lambda+2\alpha t}{\lambda} \ln\frac{(z-1)\sqrt{\lambda+2\alpha t}+\sqrt{\lambda}}{(z-1)\sqrt{\lambda+2\alpha t}} \bigg]
.
\end{align}
We plug this expression in~\eqref{psi_pref_1} and arrive at: 
\begin{align}
\displaystyle &\psi_0\bigg[(z-1)\sqrt{\frac{\lambda+2\alpha t}{\lambda}}+1\bigg]  
- \lambda \bigg[ -(z-1)\sqrt{\frac{\lambda+2\alpha t}{\lambda}}  \nonumber \\
\displaystyle &+\frac{1}{2}+(z-1)^2 \frac{\lambda+2\alpha t}{\lambda} \ln\frac{(z-1)\sqrt{\lambda+2\alpha t}+\sqrt{\lambda}}{(z-1)\sqrt{\lambda+2\alpha t}} \bigg] \nonumber \\
\displaystyle &= \psi(z,t) - (\lambda + 2 \alpha t) \bigg[ 1-z+\frac{1}{2}+(z-1)^2 \ln\frac{z}{z-1} \bigg]
.
\label{psi_temp_1}
\end{align}
Now, let us make the following simplifications: 
\eq{
\begin{cases}
\displaystyle \ln\frac{(z-1)\sqrt{\lambda+2\alpha t}+\sqrt{\lambda}}{(z-1)\sqrt{\lambda+2\alpha t}} 
= \ln \bigg[ 1+\frac{1}{z-1}\sqrt{\frac{\lambda}{\lambda+2\alpha t}} \bigg] \\ \\
\displaystyle  \ln\frac{z}{z-1}= -\ln(1-z^{-1}).
\end{cases}
}
Using these simplifications and rearranging the terms in~\eqref{psi_temp_1}, we get:
\begin{align}
\displaystyle 
&\psi(z,t)= \psi_0\bigg[(z-1)\sqrt{\frac{\lambda+2\alpha t}{\lambda}}+1\bigg] - 2\alpha t(z-1)+\alpha t \nonumber \\ 
&-2\alpha 2(z-1)^2\ln(1-z^{-1}) +(z-1) \lambda \sqrt{\frac{\lambda+2\alpha t}{\lambda}}\nonumber \\
& - \lambda(z-1)-\lambda(z-1)^2\ln(1-z^{-1})
\nonumber \\
&-(z-1)^2(\lambda+2\alpha t) \ln \bigg[ 1+\frac{1}{z-1}\sqrt{\frac{\lambda}{\lambda+2\alpha t}}\bigg]
.
\end{align}


\section{Finding the Inverse Z-transform of~\eqref{psi_temp_4} for Single Preferential Attachment} \label{app:Z_pref_single}

The expression for the generating function is as follows:
\begin{align}
\displaystyle 
\psi(z,t)&= \psi_0 \left( \frac{z-c}{1-c} \right)
 +\alpha t   -(z-1)(\lambda+2\alpha t) c \nonumber \\
&-(\lambda+2\alpha t) (z-1)^2 \ln\left(1-cz^{-1} \right).
\end{align}
Since ${c \leq 1,~\forall t}$ and $z>1$, we have $cz^{-1}<1$. So the logarithm can be Taylor-expanded.  
For  ${\ln(1-x)}$ with ${|x|<1}$ we have: 
\eq{
\ln(1-x) \sim -x-\frac{x^2}{2}-\frac{x^3}{3}-\ldots
.}
Using this, we have: 
\begin{align}
&(z-1)^2 \ln(1-cz^{-1}) \nonumber \\
&= \bigg( -cz^{-1}-\frac{c^2 z^{-2}}{2} - \frac{c^3 z^{-3}}{3} -\ldots \bigg)  \nonumber \\
&+ \bigg( -cz -\frac{c^2}{2} - \frac{c^3 z^{-1}}{3} -\frac{c^4 z^{-2}}{4} -\ldots \bigg) 
\nonumber \\
&+2\bigg( c+\frac{c^2 z^{-1}}{2}+\frac{c^3 z^{-2}}{3} + \frac{c^4 z^{-3}}{4} \ldots \bigg).
\end{align}
Using this expression for the term with the logarithm in~\eqref{psi_temp_4},  we get: 
\begin{align}
\displaystyle 
\psi(z,t)&= \psi_0 \left( \frac{z-c}{1-c} \right)
 +\alpha t   -(z-1)(\lambda+2\alpha t) c \nonumber \\
&+(\lambda+2\alpha t)  \bigg( cz^{-1}+\frac{c^2 z^{-2}}{2} + \frac{c^3 z^{-3}}{3} +\ldots \bigg)  \nonumber \\
&+ (\lambda+2\alpha t) \bigg( cz +\frac{c^2}{2} + \frac{c^3 z^{-1}}{3} +\frac{c^4 z^{-2}}{4} + \ldots \bigg) 
\nonumber \\
&-2(\lambda+2\alpha t) \bigg( c+\frac{c^2 z^{-1}}{2}+\frac{c^3 z^{-2}}{3} + \frac{c^4 z^{-3}}{4} \ldots \bigg).
\label{psi_temp_5}
\end{align}
Note that this expression seemingly embodies terms with nonegative powers of $z$. Since the sequence of degree population $N_k$ is zero for nonpositive values of $k$, the Z-transform is expected to only exhibit negative powers of $z$. Let us explicitly examine these terms and check that they do add up to zero. To do so,  let us rewrite the terms which embody nonnegative powers of $z$, which are: 
\begin{align}
&\alpha t   -(z-1)(\lambda+2\alpha t) c + (\lambda+2\alpha t) cz \nonumber \\
&+ (\lambda+2\alpha t) \frac{c^2}{2}-2(\lambda+2\alpha t) c \nonumber \\
&= \alpha t- (\lambda+2\alpha t) c + (\lambda+2\alpha t) \frac{c^2}{2}.
\end{align}
The $z^1$ term readily vanishes. Now we focus on the constant terms. Using the fact that 
\eq{
(\lambda+2\alpha t) \frac{c^2}{2}= \lambda + \alpha t -\sqrt{\lambda}\sqrt{\lambda+2\alpha t} 
,}
we have: 
\begin{align}
&\alpha t- (\lambda+2\alpha t) c + (\lambda+2\alpha t) \frac{c^2}{2} \nonumber  \\
&=- (\lambda+2\alpha t) c + \lambda+2\alpha t -\sqrt{\lambda}\sqrt{\lambda+2\alpha t} \nonumber \\
&= (\lambda+2\alpha t) \bigg(-c+1-\sqrt{\frac{\lambda}{\lambda+2\alpha t}} \bigg) \nonumber \\
&=0.
\end{align}
So these terms do cancel out. Using this simplification, ~\eqref{psi_temp_5}  becomes: 
\begin{align}
\displaystyle 
\psi(z,t)&= \psi_0 \left( \frac{z-c}{1-c} \right) \nonumber \\
&+(\lambda+2\alpha t)  \bigg( cz^{-1}+\frac{c^2 z^{-2}}{2} + \frac{c^3 z^{-3}}{3} +\ldots \bigg)  \nonumber \\
&+ (\lambda+2\alpha t) \bigg( \frac{c^3 z^{-1}}{3} +\frac{c^4 z^{-2}}{4} +\frac{c^5 z^{-3}}{5} \ldots \bigg) 
\nonumber \\
&-2(\lambda+2\alpha t) \bigg( \frac{c^2 z^{-1}}{2}+\frac{c^3 z^{-2}}{3} + \frac{c^4 z^{-3}}{4} \ldots \bigg).
\label{psi_temp_6_1}
\end{align}
which can be expressed in the following compact form: 
\begin{align}
\displaystyle 
\psi(z,t)&= \psi_0 \left( \frac{z-c}{1-c} \right) \nonumber \\
&+(\lambda+2\alpha t) \sum_{k=1}^{\infty} z^{-k} \bigg( \frac{c^k}{k}-2\frac{c^{k+1}}{k+1}+\frac{c^{k+2}}{k+2}\bigg) .
\label{psi_temp_6}
\end{align}
Now we have to take the inverse Z-transform. The sum on the right hand side is readily in the form of an expansion on $z^{-1}$. We focus on the first term on the right hand side. 
Let us denote the number of nodes in the initial graph who have degree $k$ by $N_k(0)$. Then by definition, we have: 
\eq{
\psi_0(z)=\sum_k N_k(0) z^{-k}
.}
If we change the argument of $\psi$ from $z$ to $\frac{z-c}{1-c}$, we have: 
\eq{ 
\psi_0 \left( \frac{z-c}{1-c} \right)= \sum_{\ell} N_{\ell}(0) \bigg(\frac{1-c}{z-c}\bigg)^{\ell}
.}
The inverse transform is given by: 
\begin{align}
\displaystyle &\frac{1}{2\pi i} \oint \sum_{\ell} N_{\ell}(0) \bigg(\frac{1-c}{z-c}\bigg)^{\ell} z^{k-1} dz  \nonumber \\
& =\sum_{\ell} N_{\ell}(0) \frac{(1-c)^{\ell}}{2\pi i} \oint \frac{z^{k-1}}{(z-c)^{\ell}} dz.
\label{Nk_temp_1}
\end{align}
Note that the residue of the  function $\frac{f(z)}{(z-c)^{\ell}}$ for a differentiable function $f(\cdot)$, is given by $\frac{f^{(\ell-1)}}{(\ell-1)!}$, evaluated at $z=c$. For our problem, 
$f(z)=z^{k-1}$. So we have to evaluate the $(\ell-1)$-th derivative of the function $z^{k-1}$. We have: 
\eq{
\frac{1}{(\ell-1)!}\frac{d^{(\ell-1)}}{dz^{(\ell-1)}} z^{k-1} \bigg|_{z=c}=
\begin{cases}
0~~~~~&\ell>k \\ \\
\displaystyle \frac{(k-1)!}{(\ell-1)!(k-\ell)!} c^{k-\ell} &\ell\leq k.
\end{cases}
}
Using this, we find the integrals in~\eqref{Nk_temp_1} and arrive at: 
\begin{align}
\displaystyle &\frac{1}{2\pi i} \oint \sum_{\ell} N_{\ell}(0) \bigg(\frac{1-c}{z-c}\bigg)^{\ell} z^{k-1} dz  \nonumber \\
& =\sum_{\ell}  N_{\ell}(0) (1-c)^{\ell} c^{k-\ell} \frac{(k-1)!}{(k-\ell)!}.
\label{Nk_temp_2}
\end{align}
So the inverse Z-transform of~\eqref{psi_temp_6} becomes: 
\begin{align}
\displaystyle 
N_k(t)&= \sum_{\ell}  N_{\ell}(0) (1-c)^{\ell} c^{k-\ell} \binom{k-1}{\ell-1} \nonumber \\
&+ (\lambda+2\alpha t) \bigg( \frac{c^k}{k}-2\frac{c^{k+1}}{k+1}+\frac{c^{k+2}}{k+2}\bigg) .
\label{psi_temp_7_1}
\end{align}
Replacing $\lambda$ by $N(0) \bar{k}_0$, we get:
 \begin{align}
\displaystyle 
N_k(t)&= \sum_{\ell} N_{\ell}(0) \binom{k-1}{\ell-1}   (1-c)^{\ell} c^{k-\ell}  \nonumber \\
&+ (N(0) \bar{k}_0+2\alpha t) \bigg( \frac{c^k}{k}-2\frac{c^{k+1}}{k+1}+\frac{c^{k+2}}{k+2}\bigg) .
\label{psi_temp_7}
\end{align}


\section{Solving Equation~\eqref{PDE_pref_mult} Using the Method Of Characteristics} \label{app:PDE_pref_multiple}
The differential equation we intend to solve is:
\eq{
\frac{\partial \psi}{\partial t}- \frac{\alpha \beta(z-1)}{\lambda + 2\alpha \beta t} 
\frac{\partial \psi}{\partial z}= \alpha z^{-\beta}
.}
Using the method of characteristics, we have : 
\eq{
\displaystyle \frac{dt}{1}=\frac{dz}{\big[- \frac{\alpha \beta (z-1)}{\lambda + 2\alpha \beta t}  \big]}
=\frac{d\psi}{\alpha z^{-\beta}}
.
}
The first equation yields a similar result as the previous section: 
\eq{
(z-1)^2 (\lambda + 2\alpha \beta t)= C
.
}
The second equation is the following: 
\eq{
\frac{d\psi}{\alpha z^{-\beta}}= \frac{-dz}{\alpha \beta (z-1)}\frac{C}{(z-1)^2}
.}
Let us define: 
\eq{
F(z) \stackrel{\text{def}}{=} \int^z \frac{x^{-\beta}}{(x-1)^3} dx
.}
Then we have:
\eqq{
\psi(z,t)=\frac{-C}{\beta} F(z)+\Phi(C)
.}{psi_pref_mult_1}
This can be used to determine the unknown function $\Phi(\cdot)$. For $t=0$ we have: 
\eq{
\psi_0(z)+\frac{(z-1)^2 \lambda}{\beta} F(z)= \Phi \big[ (z-1)^2 \lambda \big]
.}
From this, we find that the function $\Phi(X)$ is
\eq{
\Phi(X)=\psi_0 \bigg( \sqrt{\frac{X}{\lambda}} +1 \bigg) 
+\frac{X}{\beta} F\bigg( \sqrt{\frac{X}{\lambda}} +1 \bigg)
.}
Substituting $C$ for $X$, ~\eqref{psi_pref_mult_1} transforms to: 
\begin{align}
&\psi(z,t)=\psi_0\bigg((z-1) \sqrt{ \frac{\lambda+2\alpha \beta t}{\lambda}}+1\bigg) \nonumber \\
& -\frac{(z-1)^2 (\lambda + 2\alpha \beta t)}{\beta} \bigg[ F(z)-F\bigg((z-1) \sqrt{ \frac{\lambda+2\alpha \beta t}{\lambda}}+1\bigg) \bigg]  .
\end{align}


\section{Taking the Inverse Z-transform of~\eqref{psi_pref_mult3} for Multiple Preferential Attachment} \label{app: Z_pref_multiple}

We want to find the inverse transform of the following generating function: 
\begin{align}
&\psi(z,t)=\psi_0\bigg(\frac{z-c}{1-c}\bigg) \nonumber \\
& -\frac{(z-1)^2 (\lambda + 2\alpha \beta t)}{\beta} \bigg[ F(z)-F\bigg( \frac{z-c}{1-c} \bigg) \bigg]  .
\end{align}
First, let us find the explicit form of~$F(z)$. Note that through partial fraction expansion, we have: 
\begin{align}
&\frac{1}{x^{\beta}(x-1)^3}= \frac{\beta(\beta+1)}{2} \bigg[ \frac{1}{x-1}-\frac{1}{x} \bigg] \nonumber \\
& - \frac{\beta}{(x-1)^2}+ \frac{1}{(x-1)^3} \nonumber \\
& - \frac{1}{2} \sum_{k=2}^{\beta} (\beta-k+1)  (\beta-k+2)  z^{-k}
\end{align}
Integrating this, we obtain:
\begin{align}
&F(z)= \int^z \frac{x^{-\beta}}{(x-1)^3} dx=  \frac{\beta}{z-1} -\frac{1}{2(z-1)^2}\nonumber \\
&+\frac{\beta(\beta+1)}{2}\ln(1-z^{-1})
+ \frac{1}{2} \sum_{k=1}^{\beta-1}  \frac{(\beta-k)(\beta-k+1)}{k}z^{-k}.
\end{align}
Multiplying this expression by $(z-1)^2$ (as it appears in the generating function we want to invert),   we have:
\begin{align}
&G(z) \stackrel{\text{def}}{=} (z-1)^2 F(z)=  \nonumber \\
&\beta(z-1)-\frac{1}{2} +\frac{\beta(\beta+1)}{2}(z-1)^2 \ln(1-z^{-1})  \nonumber \\
&+ \frac{1}{2} \sum_{k=1}^{\beta-1}  \frac{(\beta-k)(\beta-k+1)}{k}(z^2-2z+1) z^{-k}.
\label{G_def_1}
\end{align}
Now let us focus on taking the inverse Z-transform of $G(z)$. Expanding the logarithm, we have: 
\begin{align}
&G(z) =  
\beta(z-1)-\frac{1}{2} \nonumber \\
&+\frac{\beta(\beta+1)}{2}(z-1)^2 \left( -z^{-1}-\frac{z^{-2}}{2}-\frac{z^{-3}}{3}-\ldots \right)   \nonumber \\
&+ \frac{1}{2} \sum_{k=1}^{\beta-1}  \frac{(\beta-k)(\beta-k+1)}{k}(z^2-2z+1) z^{-k}.
\label{G_def_2}
\end{align}
Note that the terms involving nonnegative powers f $z$ must vanish, as above. First, let us look at the $z^0$ terms in $G(z)$. The second term gives: 
\eq{
\frac{\beta(\beta+1)}{2} \left( 2-\frac{1}{2} \right) .
}
The third term in~\eqref{G_def_2}, gives two $z^0$ terms, one for ${m=1}$ and one for ${m=2}$. We have: 
\eq{
\frac{(\beta-2)(\beta-1)}{4}- \beta (\beta-1).
}
So the total $z^0$ term of $G(z)$ is: 
\eq{
-\beta-\frac{1}{2} + \frac{3 \beta(\beta+1)}{4} + \frac{(\beta-2)(\beta-1)}{4}- \beta (\beta-1).
}
One can readily check that they do add up to zero. 

Now let us examine the $z^1$ terms in $G(z)$. The second term in~\eqref{G_def_2} gives: 
\eq{
\frac{\beta(\beta+1)}{2} (-1z)
.}
The summation yields a $z^1$ term only for $m=1$. This term is: 
\eq{
\frac{\beta(\beta-1)}{2} (z).
}
Plugging these in~\eqref{G_def_2}, we find that the total $z^1$ portion of $G(z)$ is: 
\eq{
\beta z + \frac{\beta(\beta+1)}{2} (-1z) + \frac{\beta(\beta-1)}{2} (z),
}
which adds up to zero.

To take the inverse Z-transform of~\eqref{G_def_1}, let us rewrite it as follows: 
\begin{align}
&G(z) =  
\frac{\beta(\beta+1)}{-2} \sum_{k=1}^{\infty} z^{-k}\Bigg[ \frac{1}{k}-\frac{2}{k+1}+\frac{1}{k+2} \Bigg] 
\nonumber \\
&+ \frac{1}{2} \sum_{k=1}^{\beta-1}  \frac{(\beta-k)(\beta-k+1)}{k}(z^2-2z+1) z^{-k}.
\label{G_def_3}
\end{align}
It can be further simplified as follows: 
\begin{align}
&G(z) =  
\frac{\beta(\beta+1)}{-2} \sum_{k=1}^{\infty} z^{-k} \frac{2}{k(k+1)(k+2)}
\nonumber \\
 &+ \frac{1}{2} \sum_{k=1}^{\beta-1} z^{-k} \Bigg\{   
\frac{(\beta-k)(\beta-k+1)}{k} \nonumber \\
&-2  \frac{(\beta-k-1)(\beta-k)}{k+1}     
+ \frac{(\beta-k-2)(\beta-k-1)}{(k+2)}  \Bigg\} .
\label{G_def_3_2}
\end{align}
Taking the common denominator of the terms inside the summation, this expression  is simplified further and transforms to the following: 
\begin{align}
&G(z) =  
\frac{\beta(\beta+1)}{-2} \sum_{k=1}^{\infty} z^{-k} \frac{2}{k(k+1)(k+2)}
\nonumber \\
 &+ \frac{1}{2} \sum_{k=1}^{\beta-1} z^{-k} \frac{2\beta(\beta+1)}{k(k+1)(k+2)}
\label{G_def_4}
\end{align}
Since the terms are identical, $G(z)$ simplifies to the following compact form: 
\eqq{
G(z)=-\sum_{k=\beta}^{\infty} \frac{\beta(\beta+1)}{k(k+1)(k+2)}z^{-k}.
}{G_final}
Taking the inverse Z-transform is straightforward: 
\eqq{
G(z) \xrightarrow{\mathcal{Z}^{-1}} \frac{-\beta(\beta+1)}{k(k+1)(k+2)} u(k-\beta).
}{G_ZZ}
Getting back to~\eqref{psi_pref_mult3}, we can rewrite it as follows: 
\begin{align}
&\psi(z,t)=\psi_0\bigg(\frac{z-c}{1-c}\bigg) \nonumber \\
& -\frac{(\lambda + 2\alpha \beta t)}{\beta} \Bigg[ G(z)-(1-c)^2 G\bigg( \frac{z-c}{1-c} \bigg) \Bigg]  .
\label{psi_Z_final_1}
\end{align}
Note that we have: 
\eq{
(1-c)^2= \frac{\lambda}{\lambda+2\alpha \beta t},
}
so~\eqref{psi_Z_final_1} is further simplified: 
\begin{align}
&\psi(z,t)=\psi_0\bigg(\frac{z-c}{1-c}\bigg) \nonumber \\
& -\frac{(\lambda + 2\alpha \beta t)}{\beta} \Bigg[ G(z)-\frac{\lambda}{\lambda+2\alpha \beta t} G\bigg( \frac{z-c}{1-c} \bigg) \Bigg]  .
\label{psi_Z_final}
\end{align}
We have previously taken the inverse transform of the first term in~\eqref{psi_temp_7}, which was given by~\eqref{Nk_temp_2}. Using the same procedure, we know how to extract the inverse transform of a function whose argument is ${\bigg( \frac{z-c}{1-c} \bigg)}$, once the inverse transform of the function is known. That is, knowing the inverse transform of $G(z)$, we can find the inverse transform of ${G\bigg( \frac{z-c}{1-c} \bigg)}$. Let us denote the inverse transform of $G(z)$ by $g_k$. Then we have: 
\eqq{
G\bigg( \frac{z-c}{1-c} \bigg) \xrightarrow{\mathcal{Z}^{-1}} 
\sum_{\ell=1}^{k} g_{\ell} (1-c)^{\ell} c^{k-\ell} \binom{k-1}{\ell-1}. 
}{GG_ZZ}
Using this, along with~\eqref{G_ZZ}, we take the inverse Z-transform of~\eqref{psi_Z_final}. Also, let us replace $\lambda$ by $N(0) \bar{k}_0$. We obtain: 
\begin{align}
&N_k(t)= \sum_{\ell=1}^{k} N_{\ell}(0) (1-c)^{\ell} c^{k-\ell} \binom{k-1}{\ell-1} \nonumber \\
& +\frac{(N(0) \bar{k}_0 + 2\alpha \beta t)}{\beta} \frac{\beta(\beta+1)}{k(k+1)(k+2)} u(k-\beta) 
\nonumber \\
&- \frac{N(0) \bar{k}_0}{\beta}\sum_{\ell=\beta}^{k} \frac{\beta(\beta+1)}{\ell(\ell+1)(\ell+2)} (1-c)^{\ell} c^{k-\ell} \binom{k-1}{\ell-1} 
\label{Nk_pref_final_res}
\end{align}


\section{Proof of Equivalence of~\eqref{pk_pref_single} and~\eqref{pk_pref_multiple} for the Special Case of~$\beta=1$} \label{app:equivalence}
The degree distribution of single preferential attachment scheme is given by
\begin{align}
\displaystyle 
p_k(t)&= \frac{N(0)}{N(0)+\alpha t}\sum_{\ell}  n_{\ell} \binom{k-1}{\ell-1} (1-c)^{\ell} c^{k-\ell}  \nonumber \\
&+ \frac{N(0) \bar{k}_0+2\alpha t}{N(0)+\alpha t} \bigg( \frac{c^k}{k}-2\frac{c^{k+1}}{k+1}+\frac{c^{k+2}}{k+2}\bigg) ,
\label{pk1}
\end{align}
the equivalence of whom, we wish to prove, with the degree distribution of general~$\beta$-fold preferential attachment in the special case of~$\beta=1$ which is the following
\begin{align}
&p_k(t)= \frac{N(0)}{N(0)+\alpha t}\sum_{\ell}  n_{\ell} \binom{k-1}{\ell-1}  (1-c)^{\ell} c^{k-\ell} 
\nonumber \\
& +\frac{(N(0) \bar{k}_0 + 2\alpha t)}{N(0)+\alpha t} \frac{2}{k(k+1)(k+2)} 
\nonumber \\
&-\frac{2N(0) \bar{k}_0}{N(0)+\alpha t}\sum_{\ell=1}^{k} \frac{1}{\ell(\ell+1)(\ell+2)} (1-c)^{\ell} c^{k-\ell} \binom{k-1}{\ell-1}  
\label{pk2}.
\end{align}
The first sums are the same in both expressions. So is the denominator~$N(0)+\alpha t$ in every term. Hence we need to show that 
\begin{align}
&(N(0) \bar{k}_0 + 2\alpha t) \bigg( \frac{c^k}{k}-2\frac{c^{k+1}}{k+1}+\frac{c^{k+2}}{k+2}\bigg) \nonumber \\
&= (N(0) \bar{k}_0 + 2\alpha t)\frac{2}{k(k+1)(k+2)} \nonumber \\
&- 2N(0) \bar{k}_0 \sum_{\ell=1}^{k} \frac{1}{\ell(\ell+1)(\ell+2)} (1-c)^{\ell} c^{k-\ell} \binom{k-1}{\ell-1} 
\label{app_temp_1}.
\end{align}
Now let us focus on the last sum
\begin{align}
\sigma \stackrel{\text{def}}{=} \sum_{\ell=1}^{k} \frac{1}{\ell(\ell+1)(\ell+2)} (1-c)^{\ell} c^{k-\ell} \binom{k-1}{\ell-1} .
\end{align}
Expanding the binomial coefficient, we have
\al{
\sigma=  \sum_{\ell=1}^{k}  \frac{(k-1)!}{(\ell+2)! (k-\ell)!} (1-c)^{\ell} c^{k-\ell}
,}
which can be equivalently written as
\al{
\sigma=  \frac{(1-c)^{-2}}{k(k+1)(k+2)}\sum_{\ell=1}^{k}  \frac{(k+2)!}{(\ell+2)! (k-\ell)!} (1-c)^{\ell+2} c^{k-\ell}
,}
or more compactly,
\al{
\sigma=  \frac{(1-c)^{-2}}{k(k+1)(k+2)}\sum_{\ell=1}^{k} \binom{k+2}{\ell+2} (1-c)^{\ell+2} c^{k-\ell}
.}
Note that if this sum had commenced at~$\ell=-2$, it would be the binomial expansion of
\eq{
[(1-c)+(c)]^{k+2}
,}
which is identical to unity. So we can write this sum as unity minus the three missing terms: 
\al{
&\sigma=  \frac{1}{k(k+1)(k+2)} \overbrace{\bigg[ \frac{N(0) \bar{k}_0 + 2\alpha t}{N(0) \bar{k}_0}\bigg]}^{\textnormal{replaced for } (1-c)^{-2}} \Bigg[ 1- c^{k+2}
\nonumber \\
&- (k+2) (1-c)c^{k+1}- \frac{(k+1)(k+2)}{2} (1-c)^2 c^{k}  \Bigg] 
.}
Plugging  this expression in~\eqref{app_temp_1}, we get
\begin{align}
&(N(0) \bar{k}_0 + 2\alpha t) \bigg( \frac{c^k}{k}-2\frac{c^{k+1}}{k+1}+\frac{c^{k+2}}{k+2}\bigg) \nonumber \\
&= (N(0) \bar{k}_0 + 2\alpha t)\frac{2}{k(k+1)(k+2)} \nonumber \\
&- \frac{2 (N(0) \bar{k}_0 + 2\alpha t) }{k(k+1)(k+2)}
 \bigg[ 1- c^{k+2} \nonumber \\
&- (k+2) (1-c)c^{k+1}- \frac{(k+1)(k+2)}{2} (1-c)^2 c^{k}  \bigg]
\label{app_temp_2}.
\end{align}
Canceling out the term~$(N(0) \bar{k}_0 + 2\alpha t)$, and then multiplying both sides of the equality by~$k(k+1)(k+2)$, we get 
\begin{align}
& c^k(k+1)(k+2)-2c^{k+1} k(k+2)+c^{k+2}k(k+1) 
 \nonumber \\
&= 2 -  2+2 c^{k+2} 
 +2 (k+2) (1-c)c^{k+1}\nonumber \\
&+ (k+1)(k+2)  (1-c)^2 c^{k} 
\label{app_temp_3}.
\end{align}
 Rearranging the terms on the right hand side in powers of $c$, we have: 
\begin{align}
& c^k(k+1)(k+2)-2c^{k+1} k(k+2)+c^{k+2} k(k+1)   \nonumber \\
&=c^{k+2}  \bigg[ 2-2(k+2)+(k+1)(k+2) \bigg] \nonumber \\
 &+ c^{k+1} \bigg[  2(k+2)-2(k+1)(k+2) \bigg]  \nonumber \\
&+ c^k \bigg[   (k+1)(k+2)  \bigg] 
\label{app_temp_3}.
\end{align}
It remains to show that the coefficients of~$c^{k+1}$ and ~$c^{k+2}$ on the left hand side are equal to those on the right hand side. For~$c^{k+1}$, note that
\eq{
 2(k+2)-2(k+1)(k+2)=2(k+2)(1-k-1)=-2k(k+2)
,}
which is also identical to the pertinent coefficient on the left and side. Finally, for the coefficient of~$c^{k+2}$ on the right hand side, we have
\begin{align}
&2-2(k+2)+(k+1)(k+2) \nonumber \\
&=2-2k-4+k^2+3k+2=k^2+k=k(k+1)
,
\end{align}
which is the same as the coefficient of~$c^{k+2}$ on the left hand side. Hence~\eqref{pk1} and~\eqref{pk2} are identical.

%

\end{document}